\documentclass[10pt,amsmath,amssymb,aps,nofootinbib,notitlepage,prd,superscriptaddress,twocolumn]{revtex4-1}

\usepackage[T1]{fontenc}
\usepackage[usenames,dvipsnames]{xcolor}
\usepackage{aas_macros,graphicx,hyperref,mathrsfs,orcidlink,subfigure}
\hypersetup{colorlinks=true,citecolor=RoyalBlue,linkcolor=RoyalBlue,urlcolor=RoyalBlue}

\let\originalleft\left
\let\originalright\right
\renewcommand{\left}{\mathopen{}\mathclose\bgroup\originalleft}
\renewcommand{\right}{\aftergroup\egroup\originalright}

\newcommand{\ab}[1]{\left|#1\right|}
\newcommand{\av}[1]{\left\langle#1\right\rangle}
\newcommand{\br}[1]{\left[#1\right]}

\newcommand{\pa}[1]{\left(#1\right)}
\newcommand{\ed}{\mathop{}\!\mathrm{d}}
\DeclareMathOperator\arcsinh{arcsinh}

\begin{document}

\title{\normalsize{Explanation for the absence of secondary peaks in black hole light curve autocorrelations}}

\author{Alejandro C\'ardenas-Avenda\~no\,\orcidlink{0000-0001-9528-1826}} 
\email{cardenas-avendano@princeton.edu}
\affiliation{Princeton Gravity Initiative, Princeton University, Princeton, New Jersey 08544, USA}
\affiliation{Department of Physics, Princeton University, Princeton, New Jersey 08544, USA}

\author{Charles Gammie\,\orcidlink{0000-0001-7451-8935}}
\email{gammie@illinois.edu}
\affiliation{Astronomy, Physics, NCSA, and ICASU, University of Illinois, Urbana, Illinois 61801, USA}

\author{Alexandru Lupsasca\,\orcidlink{0000-0002-1559-6965}}
\email{alexandru.v.lupsasca@vanderbilt.edu}
\affiliation{Department of Physics \& Astronomy, Vanderbilt University, Nashville, Tennessee 37212, USA}

\begin{abstract}
The observed radiation from hot gas accreting onto a black hole depends on both the details of the flow and the spacetime geometry.
The lensing behavior of a black hole produces a distinctive pattern of autocorrelations within its photon ring that encodes its mass, spin, and inclination.
In particular, the time autocorrelation of the light curve is expected to display a series of peaks produced by light echoes of the source, with each peak delayed by the characteristic time lapse $\tau$ between light echoes.
However, such peaks are absent from the light curves of observed black holes.
Here, we develop an analytical model for such light curves that demonstrates how, even though light echoes always exist in the signal, they do not produce autocorrelation peaks if the characteristic correlation timescale $\lambda_0$ of the source is greater than $\tau$.
We validate our model against simulated light curves of a stochastic accretion model ray traced with a general-relativistic code, and then fit the model to an observed light curve for Sgr\,A$^*$.
We infer that $\lambda_0>\tau$, providing an explanation for the absence of light echoes in the time autocorrelations of Sgr\,A$^*$ light curves.
Our results highlight the importance for black hole parameter inference of spatially resolving the photon ring via future space-based interferometry.
\end{abstract}

\maketitle

\textit{Introduction.}---%
Light emitted from hot gases accreting onto black holes has been observed for decades across the electromagnetic spectrum \cite{Schmidt1963,Bowyer1965,Narayan2023}.
This radiation depends on both the details of the astrophysical sources and the spacetime geometry around the black holes.

More precisely, a single source around a black hole can produce multiple images arising from photons that circumnavigate the event horizon a different number of times on their way to the observer.
These mirror images are lensed into a distinctive ``photon ring'' that represents the stamp imprinted on a black hole image by its strong gravity \cite{Luminet1979,Gralla2019,JohnsonLupsasca2020,Lupsasca2024}, and which tracks a ``critical curve'' \cite{Bardeen1973}.

Successive images appearing within the photon ring are increasingly demagnified, rotated, and time-delayed.
In the simplest case of an equatorial source viewed by a distant observer on the black hole spin axis, these images accumulate near the critical curve and may be labeled by the number of polar half-orbits that the corresponding photons execute around the black hole before reaching the observer.
If the $n^\text{th}$ image of a point source appears at a time $t_n$, at an angle $\varphi_n$ around the critical curve and at a perpendicular distance $d_n$ from it, then one can analytically prove \cite{GrallaLupsasca2020a} that the next image will appear at
\begin{align}
    \label{eq:Lensing}
    t_{n+1}\approx t_n+\tau,\quad
    \varphi_{n+1}\approx\varphi_n+\delta,\quad
    d_{n+1}\approx e^{-\gamma}d_n,
\end{align}
where the ``critical parameters'' $\tau$, $\delta$, and $\gamma$---controlling the time delay, rotation and demagnification of strongly lensed images, respectively---are known functions of the black hole mass and spin~\cite{Teo2003,GrallaLupsasca2020a}. 
In particular, $\tau\approx16M$ for most values of the spin, where $M$ denotes the black hole mass and we work in geometric units with $G=c=1$.

Because of the lensing behavior \eqref{eq:Lensing} of the black hole, the autocorrelation of its photon ring image intensity must display a distinctive multi-peaked structure, with the heights and locations of successive peaks respectively demagnified by $e^{-\gamma}$ and shifted in the spatio-temporal correlation plane $(\Delta t,\Delta\phi)$ by $(\tau,\delta)$
\cite{Hadar2021} (Fig.~1 therein).

The recent horizon-scale images taken by the Event Horizon Telescope (EHT) of the supermassive black holes M87$^*$ \cite{EHT2019a,EHT2024} and Sgr\,A$^*$ (the one at the center of our Galaxy) \cite{EHT2022a} are unable to resolve their photon rings.
Vigorous efforts to extend the EHT array to space are now underway \cite{Gurvits2022,Kurczynski2022,Kudriashov2021}, and future observations using very-long-baseline interferometry (VLBI) to space
could achieve the resolution needed to measure these rings and their predicted (ring-averaged) autocorrelations \cite{Hadar2021}.

In the meantime, one can already access the black hole light curves observed over many frequencies and for many sources, including active galactic nuclei, X-ray binaries, and gamma-ray bursts \cite{Ulrich1997,Remillard2006,Berger2014}.
Such light curves may be regarded as (single-pixel) ``images'' that are completely spatially averaged (over both radius and angle).
Based on the preceding discussion, one would expect the temporal autocorrelation of many of these light curves to display multiple peaks, with each successive one demagnified by $e^{-\gamma}$ and delayed in time by the characteristic interval $\tau$ between light echoes \cite{Fukumura2008,Moriyama2019,Chesler2020,Wong2021,Hadar2021,Hadar2023}.
However, such time autocorrelations have never been detected.

In particular, an analysis of a decade of  230\,GHz light curves of Sgr\,A$^*$ reported a characteristic autocorrelation timescale of $8_{-4}^{+3}$ hours at $95\%$ confidence, present down to at least a few Schwarzschild radii \cite{Dexter2014}.
Assuming a mass of $4.3\times10^6M_{\odot}$, this value corresponds in geometric units to a timescale of $1361_{-680}^{+510}M$, which is significantly higher than the expected light echo time delay $\tau\approx16M$.
An analysis of newer data collected during the 2017 EHT observation campaign of Sgr\,A$^*$ reported a characteristic autocorrelation timescale of $\sim1$ hour, or $\sim 170M$~\cite{Wielgus2022}.

Despite being significantly lower, this timescale is still much longer than the expected light echo time delay of $\tau\approx5$ minutes.
This raises the obvious question:
\begin{center}
    \textit{Where are the light echo autocorrelation peaks?}
\end{center}

In this letter, we revisit the theoretical expectations for the time autocorrelation of a black hole light curve and argue that the peaks caused by lensed images of the main emission are only present if the characteristic timescale $\lambda_0$ of temporal correlations in the source is much shorter than the light echo time delay $\tau$.
By contrast, if $\lambda_0\gtrsim\tau$, then these maxima ought to be absent, even if the lensed images are present and contribute flux to the light curve.

To support this claim, we derive an analytical model for the light curve of a black hole that is surrounded by an equatorial source observed ``face-on'' (that is, at a small inclination $\theta_{\rm o}$ from the spin axis).
We then argue that the model continues to hold provided the parameter $a_*\sin{\theta_{\rm o}}$ remains small, where $a_*=J/M^2\in[-1,1]$ denotes the black hole spin and $J$ its angular momentum.

We validate our model against simulated light curves of a stochastic accretion model that we ray trace using a general-relativistic code, and then we fit the model to an observed light curve for Sgr\,A$^*$.
We infer that $\lambda_0>\tau$, providing an explanation for the absence of secondary peaks in the time autocorrelations of Sgr\,A$^*$ light curves.

\textit{Theoretical expectations.}---%
Consider a polar observer ($\theta_{\rm o}=0^\circ$) of equatorial emission around the black hole.
We decompose the full image into layers labeled by $n$,
\begin{align}
    I(t_{\rm o},\alpha,\beta)=\sum_{n=0}^\infty I_n(t_{\rm o},\alpha,\beta),
\end{align}
where the $n^\text{th}$ layer corresponds to the image of the source produced by photons that travel $n$ half-orbits around the black hole, described in Cartesian coordinates $(\alpha,\beta)$ on the image plane at observation time $t_{\rm o}$ \cite{Bardeen1973}.
It is sometimes more convenient to use a polar angle $\varphi$ and perpendicular distance $d$ from the critical curve as image coordinates \cite{GrallaLupsasca2020a}.
It follows from the lensing equations \eqref{eq:Lensing} that
\begin{align}
    \label{eq:ImageLensing}
    I_n(t_{\rm o},\varphi,d)\approx I_{n-1}\pa{t_{\rm o}-\tau,\varphi-\delta,e^\gamma d},
\end{align}
up to small corrections in $1/n$ that are already negligible for $n\gtrsim2$.
Each image layer has a flux (``light curve'')
\begin{align}
    \mathscr{L}_n(t_{\rm o})=\int I_n(t_{\rm o},\alpha,\beta)\ed\alpha\ed\beta.
\end{align}
The lensing relation \eqref{eq:ImageLensing} implies that
\begin{align}
    \label{eq:LightCurveLensing}
    \mathscr{L}_n(t_{\rm o})\approx e^{-\gamma}\mathscr{L}_{n-1}(t_{\rm o}-\tau).
\end{align}
Hence, the total observed light curve is approximately
\begin{align}
    \label{eq:LightCurve}
    \mathscr{L}(t_{\rm o})=\sum_{n=0}^\infty\mathscr{L}_n(t_{\rm o})
    \approx\sum_{n=0}^\infty e^{-n\gamma}\mathscr{L}_0(t_{\rm o}-n\tau).
\end{align}
As expected, it consists of a superposition of multiple copies of the light curve $\mathscr{L}_0(t_{\rm o})$ of the direct emission.
Each copy carries $e^{-\gamma}$ less flux and is time-delayed by $\tau$ relative to its predecessor.
If the source is stationary, then the covariance $C_0(\Delta t)=\av{\mathscr{L}_0(t_{\rm o})\mathscr{L}_0(t_{\rm o}+\Delta t)}$ of the light curve for the direct emission is time-translation invariant, and the full light curve covariance at lag $\Delta t$ is\footnote{We use $\av{\mathscr{L}_i(t)\mathscr{L}_j(t+\Delta t)}\approx\av{\mathscr{L}_0(t-i\tau)\mathscr{L}_0(t+\Delta t-j\tau)}$ from Eq.~\eqref{eq:LightCurveLensing}, and then shift $(i,j)$ to $(m,2s)=(i-j,i+j-|m|)$.}
\begin{align}
    \label{eq:LightCurveCorrelationAux}
    C(\Delta t)&=\av{\mathscr{L}(t_{\rm o})\mathscr{L}(t_{\rm o}+\Delta t)}\\
    &=\sum_{i=0}^\infty\sum_{j=0}^\infty e^{-(i+j)\gamma}\av{\mathscr{L}_i(t)\mathscr{L}_j(t+\Delta t)}\\
    &\approx\sum_{i=0}^\infty\sum_{j=0}^\infty e^{-(i+j)\gamma}C_0\pa{\Delta t+(i-j)\tau}\\
    \label{eq:LightCurveCorrelation}
    &=\frac{1}{1-e^{-2\gamma}}\sum_{m=-\infty}^\infty e^{-|m|\gamma}C_0\pa{\Delta t+m\tau},
\end{align}
which is also stationary.
This expression agrees with the conclusions arrived at by different means in Ref.~\cite{Hadar2021} and describes a train of correlation peaks separated in time by $\tau$ and exponentially decreasing in height by $e^{-\gamma}$.

As argued in Sec.~7 of Ref.~\cite{Hadar2021}, even though the lens equations \eqref{eq:Lensing} and \eqref{eq:ImageLensing} are modified for $\theta_{\rm o}>0$, the formula \eqref{eq:LightCurveCorrelation} nevertheless remains exact to leading (linear) order in $a_*\sin{\theta_{\rm o}}$, with the first correction coming in only at subleading (quadratic) order.
Thus, the only relevant time delay at small inclinations is the orbital half-period $\tau(\tilde{r}_0)$ of the null geodesics trapped at the radius $\tilde{r}_0$ where bound photons have vanishing spin angular momentum.
Likewise, the Lyapunov exponent $\gamma(\tilde{r}_0)$ governing their orbital instability fully controls the demagnification \cite{GrallaLupsasca2020a}.

For moderate inclinations, the only meaningful change one can expect is that strongly lensed photons may skirt a range of bound orbits at different radii $\tilde{r}$ in the ``photon shell'' of trapped null geodesics \cite{JohnsonLupsasca2020}.
The time delay $\tau(\tilde{r})$ and Lyapunov exponent $\gamma(\tilde{r})$ are functions of this orbital radius, so one expects a smearing of Eq.~\eqref{eq:LightCurveCorrelation} over a range of $\tilde{r}$, which remains quite narrow up to $\theta_{\rm o}\lesssim45^\circ$.
At $\tilde{r}_0$, $\tau\approx16M$ for all $a_*$ as $e^{-\gamma}$ ranges from $e^{-\pi}\approx4\%$ to $10\%$.

Therefore, we take the expression \eqref{eq:LightCurveCorrelation} (with $\tau$ and $\gamma$ always evaluated near $\tilde{r}_0$) as our general analytical model for the covariance of the light curve.
Since $e^{-\gamma}\lesssim10\%$, we expand the time autocorrelation of the light curve as
\begin{align}
    \label{eq:Autocorrelation}
    \mathscr{C}(\Delta t)=\frac{C(\Delta t)}{C(0)}
    &\approx\mathscr{C}_0(\Delta t)\br{1-2e^{-\gamma}\mathscr{C}_0(\tau)}\\
    &\phantom{=}+e^{-\gamma}\br{\mathscr{C}_0(\Delta t+\tau)+\mathscr{C}_0(\Delta t-\tau)},\notag
\end{align}
where $\mathscr{C}_0(\Delta t)=C_0(\Delta t)/C_0(0)$ is the direct light curve time autocorrelation, and we have suppressed $\mathcal{O}\pa{e^{-n\gamma}}$ terms with $n\ge2$, incurring only negligible errors $\lesssim1\%$.

The prediction \eqref{eq:Autocorrelation} is our main theoretical result.
It is analytically well-motivated and we numerically validate it in the next section.
Our key point is that when the direct emission has a characteristic correlation timescale $\lambda_0\gtrsim\tau$, the expected peaks at $\Delta t\approx\tau$ in Eq.~\eqref{eq:Autocorrelation} vanish.
\filbreak

More precisely, Eq.~\eqref{eq:LightCurveCorrelation} predicts correlation peaks at regular intervals $\Delta t\sim m\tau$.
We now focus on the peak at $\Delta t\sim\tau$, expected to be produced by lensed images of the source whose photon half-orbit numbers differ by 1.
Here, it is important to distinguish between two regimes.

The typical time autocorrelation $\mathscr{C}_0(\Delta t)$ for a source with temporal correlations on a characteristic timescale $\lambda_0$ is a monotonically decreasing function, which starts at $\mathscr{C}_0(0)=1$ (by definition), slowly drops until $\Delta t\sim\lambda_0$, and then decays exponentially for $\Delta t\gtrsim\lambda_0$.

If $\tau\gg\lambda_0$, then $\mathscr{C}_0(\tau)\approx\mathscr{C}_0(\Delta t+\tau)\approx0$, and thus Eq.~\eqref{eq:Autocorrelation} reduces to the simpler form
\begin{align}
\label{eq:Case1}
    \mathscr{C}(\Delta t)\approx\mathscr{C}_0(\Delta t)+e^{-\gamma}\mathscr{C}_0(\Delta t-\tau),
\end{align}
describing a main peak at $\Delta t=0$ (corresponding to the perfect autocorrelation of the signal with itself), followed by the expected secondary peak at $\Delta t=\tau$.

By contrast, if $\tau\ll\lambda_0$, then $\mathscr{C}_0(\tau)\approx1$ and moreover, $\mathscr{C}_0(\Delta t\pm\tau)\approx\mathscr{C}_0(\Delta t)$ for lags $\tau\sim\lambda_0$, so Eq.~\eqref{eq:Autocorrelation} predicts
\begin{align}
    \label{eq:Explanation}
    \mathscr{C}(\Delta t)\stackrel{\Delta t\sim\tau}{\approx}\mathscr{C}_0(\Delta t).
\end{align}
Hence, there should be no secondary peak in this regime.

As reported in the Supplemental Material (SM)~\cite{SM}, we observe these two behaviors in our numerical simulations.
In the intermediate regime $\tau\sim\lambda_0$, we find that as $\tau$ increases, the secondary peak changes from a ``bump'' to an ``excess'' and usually disappears well before $\tau\approx\lambda_0$.

State-of-the-art simulations using general-relativistic magnetohydronamics (GRMHD) typically find accretion flows that circulate around the black hole at very slightly sub-Keplerian velocities $\Omega\approx\xi\Omega_{\rm K}$, with sub-Keplerianity $\xi\lesssim1$ \cite{Porth2021,Conroy2023}.
There is also experimental evidence for such behavior from observations with GRAVITY \cite{GRAVITY2024}.

If the characteristic timescale of correlations tracks the orbital period of the circularized flow, so that $\lambda_0\sim2\pi/\Omega$, then $\lambda\gg\tau$ everywhere in the gas.
In light of Eq.~\eqref{eq:Explanation}, this provides an explanation for the absence of secondary peaks in observed black hole light curve autocorrelations.

\textit{Applications.}---%
We apply our model \eqref{eq:Autocorrelation} for the time autocorrelation of light curves to (i) synthetic data and (ii) a light curve of Sgr\,A$^*$ from the 2017 EHT campaign.

For (i), we use \texttt{inoisy}~\cite{Lee2021} to simulate an equatorial source with stochastic fluctuations and \texttt{AART}~\cite{CardenasAvendano2022} to ray trace its relativistic images.
More exactly, we use \texttt{inoisy} to generate realizations of Gaussian random fields (GRF) with a Mat\'ern covariance, which serve as a proxy for a hot gas that surrounds the black hole and fluctuates with a prescribed correlation structure.
We then ray trace these realizations using \texttt{AART}, a code that exploits the integrability of light propagation in the Kerr spacetime to efficiently produce high-resolution black hole movies.

This semi-analytical approach to producing black hole movies with an analytically known covariance function is arguably the ``optimal setup'' for extracting echoes from light curves, since this method gives us complete control over all correlation scales and black hole parameters.
See the SM for the details of the implementation~\cite{SM}.

\begin{figure*}
    \centering
    \subfigure{\includegraphics[width=\textwidth]{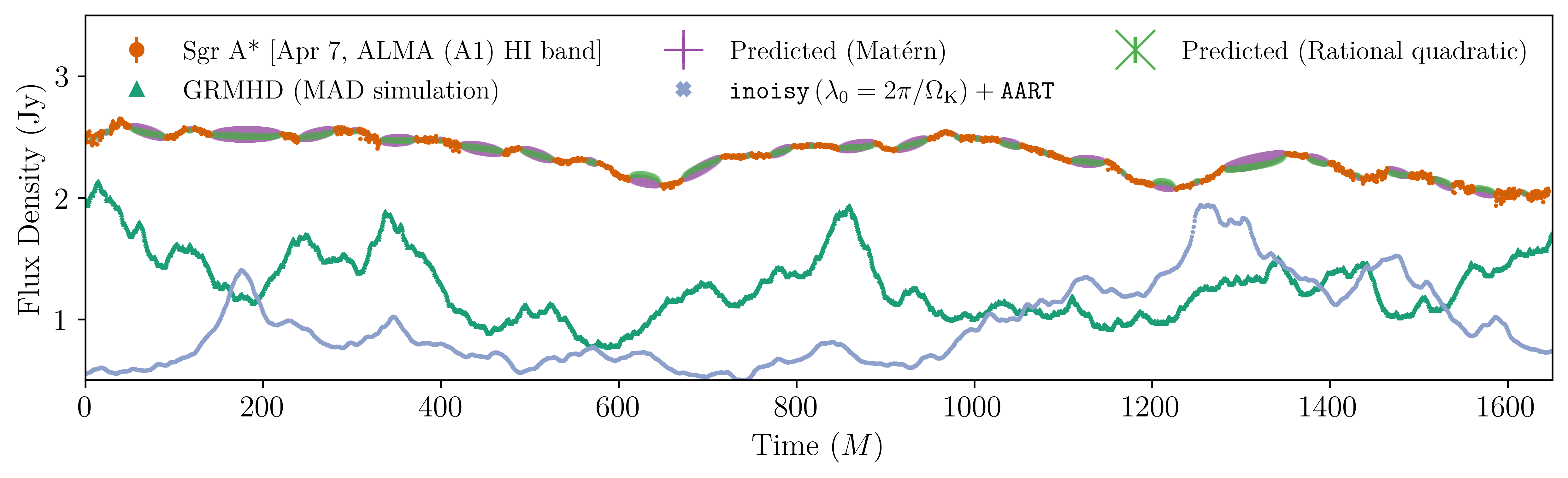}}
    \subfigure{\includegraphics[width=\columnwidth]{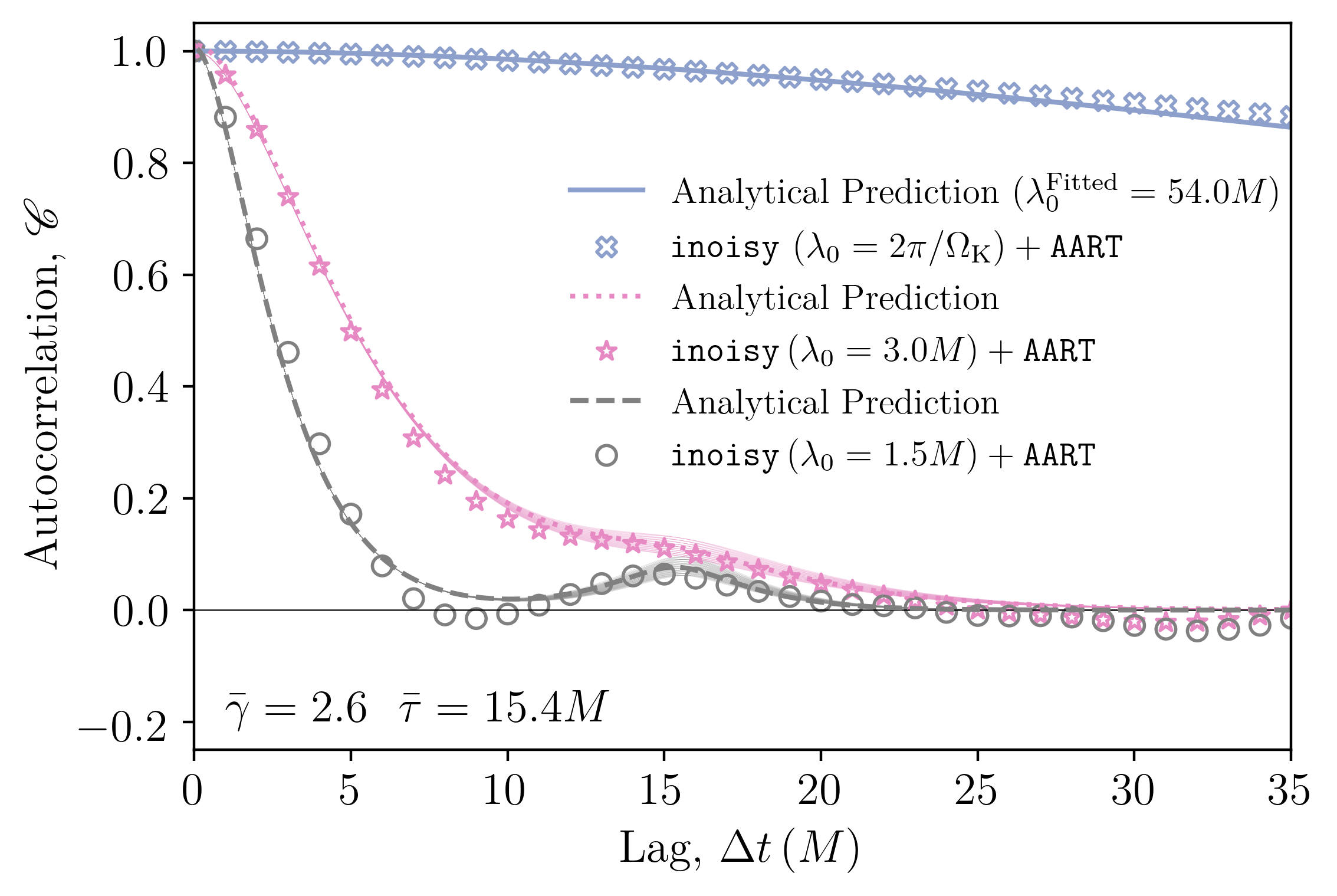}}
    \subfigure{\includegraphics[width=\columnwidth]{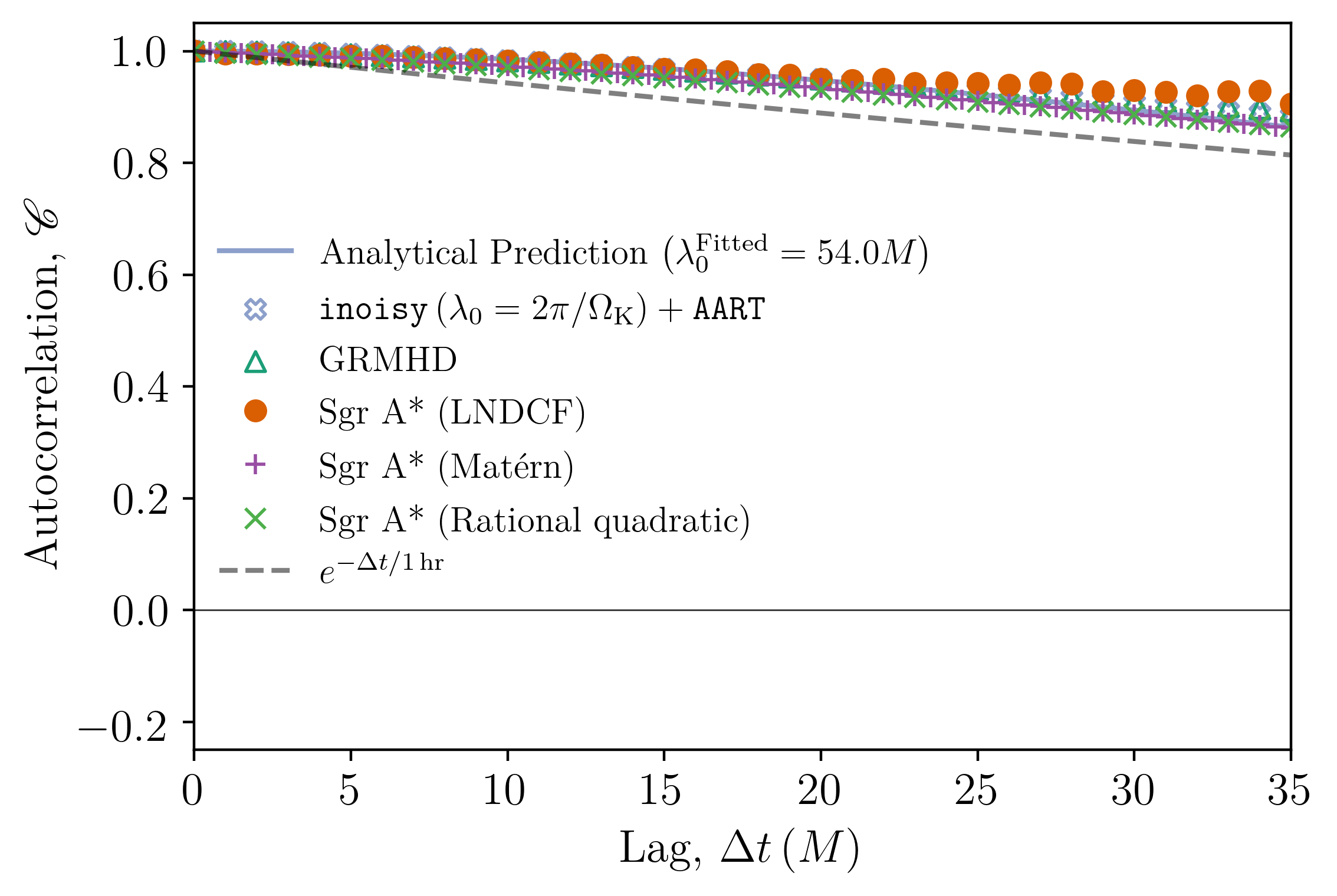}} 
    \caption{\textbf{Top panel:} Light curves from an observation of Sgr\,A$^*$ with ALMA~\cite{Wielgus2022}, from a GRMHD simulation~\cite{Wong2024}, and from an \texttt{inoisy} simulation with $\lambda_0=2\pi/\Omega_{\rm K}$ ray traced with \texttt{AART}.
    We assumed a mass $M=4.3\times10^6M_{\odot}$ for Sgr\,A$^*$ to convert time to units of $M$.
    \textbf{Bottom left:} The autocorrelation of synthetic light curves produced with \texttt{inoisy} using different correlation times and ray traced with \texttt{AART} assuming a black hole with spin $a_*=94\%$ observed from an inclination $\theta_{\rm o}=20^\circ$.
    For this geometry, the Lyapunov exponent $\gamma(\tilde{r})$ ranges over $[2.34,2.79]$ and the time delay $\tau(\tilde{r})/M$ ranges over $[15.20,15.85]$, where the lower and upper bounds correspond to the innermost radius $\tilde{r}_-$ and outermost radius $\tilde{r}_+$, respectively, within the photon shell.
    We have used their mean values (written in the plots) for the analytical predictions presented with lines.
    The transparent lines represent all the possible predictions when using the whole set of values of $\gamma$ and $\tau$.
    The grey and pink open circles correspond to underlying correlation timescales $\lambda_0=1.5M$ and $\lambda_0=3.0M$, respectively.
    The blue crosses correspond to the autocorrelation of the light curve shown in the top panel computed from an \texttt{inoisy} and \texttt{AART} simulation.
    \textbf{Bottom right:} The autocorrelations of the light curves presented in the top figure.
    For the observed light curve of Sgr\,A$^*$, we computed the autocorrelation using the LNDCF algorithm directly (orange dots), and after interpolating it with two different kernels for a Gaussian process regression (purple pluses, for Mat\'ern, and blue crosses, for rational quadratic).
    The solid line corresponds to the best-fit to Eq.~\eqref{eq:AutocorrelationHomogenenous} presented in the bottom-left panel.
    For comparison, as in Ref.~\cite{Wielgus2022}, we include as a dashed line an exponential decay with a 1 hour timescale.
    From these results, we infer that $\lambda_0>\tau$, providing an explanation for the absence of light echoes in the time autocorrelation of Sgr\,A$^*$ light curves.}
    \label{fig:All}
\end{figure*}

We consider a Kerr black hole with spin $a_*=94\%$ observed from an inclination $\theta_{\rm o}=20^\circ$, and sources with different characteristic timescales $\lambda_0$.
We compare their resulting autocorrelations $\mathscr{C}(\Delta t)$ to the prediction \eqref{eq:Autocorrelation}.
The blue line in Fig.~\ref{fig:All} (top panel) shows the light curve \eqref{eq:LightCurve} corresponding to a Keplerian flow with $\lambda_0=2\pi/\Omega_{\rm K}$, i.e., proportional to $r_{\rm s}^{3/2}$, with $r_{\rm s}$ the equatorial radius in the source.
For this position-dependent correlation time, the resulting GRF is inhomogeneous and anisotropic (see SM for an example snapshot~\cite{SM}).
The blue open crosses in Fig.~\ref{fig:All} (bottom-left panel) plot the corresponding time autocorrelation, which does not present a secondary peak at $\Delta t\sim\tau$, consistent with Eq.~\eqref{eq:Explanation} and its implications.

We stress that the absence of correlation peaks in these simulations cannot be attributed to limitations in the computation of the light curves (as we have full control of the simulation resolution and sampling rate) or in their analysis (we also applied high-pass filters and computed derivatives to search for concavity changes, to no avail).

If the source has constant correlation scales, then the light curve $\mathscr{L}_0(t_{\rm o})$ of the direct image $I_0(t_{\rm o},\alpha,\beta)$ is a Mat\'ern field with $d=1$, $\nu=3/2$ and correlation length $\lambda_0$~\cite{SM}.
Its autocorrelation is
\begin{align}
    \label{eq:AutocorrelationHomogenenous}
    \mathscr{C}_0(\Delta t)=\pa{1+\frac{\ab{\Delta t}}{\lambda_0}}\exp\pa{-\frac{\ab{\Delta t}}{\lambda_0}}.
\end{align}
We emphasize that this is not a numerical fit to \texttt{inoisy} simulations, but rather the analytical formula describing the autocorrelation of the underlying stochastic model.
This autocorrelation is also in agreement with our ray traced light curves, even when the observer is inclined and the black hole rotates rapidly, as shown in Fig.~\ref{fig:All} (bottom-left panel) for two different values of $\lambda_0$.
The case with $\lambda_0=1.5M\ll\tau$ displays a clear secondary peak (or ``bump'') while the case with $\lambda_0=3.0M<\tau$ leads only to a milder ``excess'' in the autocorrelation.

When the correlation timescale $\lambda_0$ varies across the source, we cannot provide an analytical expression for the resulting autocorrelation like the one in Eq.~\eqref{eq:AutocorrelationHomogenenous}.
We can, however, use Eq.~\eqref{eq:AutocorrelationHomogenenous} to derive an effective $\lambda_0$ by fitting it to the autocorrelation data.
For the Keplerian flow with $\lambda_0=2\pi/\Omega_{\rm K}$, we obtain a good fit with an effective $\lambda_0=54.0M$, as shown with the solid blue line in Fig.~\ref{fig:All}. 

For (ii), we compute the autocorrelation of an observed intensity light curve for Sgr\,A$^*$ and call upon the intuition built from our model to interpret the results.
Specifically, we use the April 7th, 2017 data from the $229.1$\,GHz (HI) ALMA (A1) band~\cite{Wielgus2022}.
In Fig.~\ref{fig:All} (top panel), we show the observed Sgr\,A$^*$ light curve (orange points), as well as a simulated light curve computed from an M87$^*$-like GRMHD simulation (green triangles) of a magnetically arrested disk (MAD) with $r_{\rm high}=40$ around a black hole with $a_*=85\%$ and $\theta_{\rm o}=163^\circ$~\cite{Wong2024}; see Refs.~\cite{Wong2022,Wong2024} for more details.
To obtain an autocorrelation from the observed Sgr\,A$^*$ light curve, we must account for its gaps.
As in Ref.~\cite{Wielgus2022}, we compute a locally normalized discrete correlation function (LNDCF)~\cite{Lehar1992,Edelson1988} and use Gaussian process regression (GPR)~\cite{Rasmussen2006} to interpolate the data.
This then allows us to apply the same procedure used for the synthetic light curves.
See the SM for details of these implementations~\cite{SM}.
The resulting autocorrelations are similar regardless of the method used to obtain them.

Although the GRMHD and \texttt{inoisy}+\texttt{AART} light curves in Fig.~\ref{fig:All} (top panel) are produced from very different models, their time autocorrelations (bottom-right panel) are remarkably similar to each other and also to the ones computed from the observed light curve of Sgr\,A$^*$.

Consistent with Ref.~\cite{Wielgus2022}, Fig.~\ref{fig:All} (bottom-right panel) shows no clear signs of a correlation peak at lag $\Delta t\sim\tau$ that could be interpreted as an effect of lensing by Sgr\,A$^*$.

We plot the complete autocorrelation in the SM~\cite{SM}.
Since $\mathscr{C}(\tau)\approx1$, the plot strongly suggests that $\lambda^{\rm{Sgr\,A^*}}_0>\tau$: that is, the characteristic timescale of correlations in the plasma around Sgr\,A$^*$ appears to exceed the time delay between light echoes.
Hence, in accordance with Eq.~\eqref{eq:Explanation}, we should not expect to see secondary correlation peaks, explaining their absence from observations.

In passing, we note that the Sgr\,A$^*$ autocorrelation is well-approximated by the simulated autocorrelation \eqref{eq:AutocorrelationHomogenenous} of a Keplerian flow with effective $\lambda_0=54.0M>\tau$ (solid blue lines in the bottom panels of Fig.~\ref{fig:All}).

\textit{Discussion}---%
We have developed a simple analytical model for the autocorrelation of black hole light curves that offers insight into some of the challenges involved in separating the effects of the plasma from those of the spacetime geometry.
When applied to an observed light curve of Sgr\,A$^*$, our model suggests that the temporal correlations inherent in its surrounding plasma suppress the autocorrelation peaks expected from lensing around the black hole, explaining the absence of such signatures.

These results indicate that the inference of black hole parameters from strong lensing effects will be difficult via light curve autocorrelations alone, and likely require future space-VLBI observations that spatially resolve the photon ring.
Planning for such observations is underway.

We thank Maciek Wielgus and George Wong for their valuable comments, and for providing the data for the Sgr\,A$^*$ and GRMHD light curves, respectively.
We also thank Neal Dalal, Suvendu Giri, Lennox Keeble, Aviad Levis, Leo Stein and Sam Gralla for helpful discussions.
CG and AL also thank the Aspen Center for Physics, which is supported by NSF grant PHY-2210452.
ACA acknowledges support from the Simons Foundation.
AL is supported by NSF grants AST-2307888 and PHY-2340457.
CG was supported in part by the IBM Einstein Fellow Fund at the Institute for Advanced Study.
Some of the simulations presented in this work were performed on computational resources managed and supported by Princeton Research Computing, a consortium of groups including the Princeton Institute for Computational Science and Engineering (PICSciE) and the Office of Information Technology's High Performance Computing Center and Visualization Laboratory at Princeton University.

\appendix

\onecolumngrid
\section*{}
\clearpage

\section*{Supplemental Material}

\section{Variation of the photon ring critical parameters with black hole spin and inclination}
\label{app:CriticalParameters}

\begin{figure}[h]
    \centering
	\includegraphics[width=0.4\columnwidth]{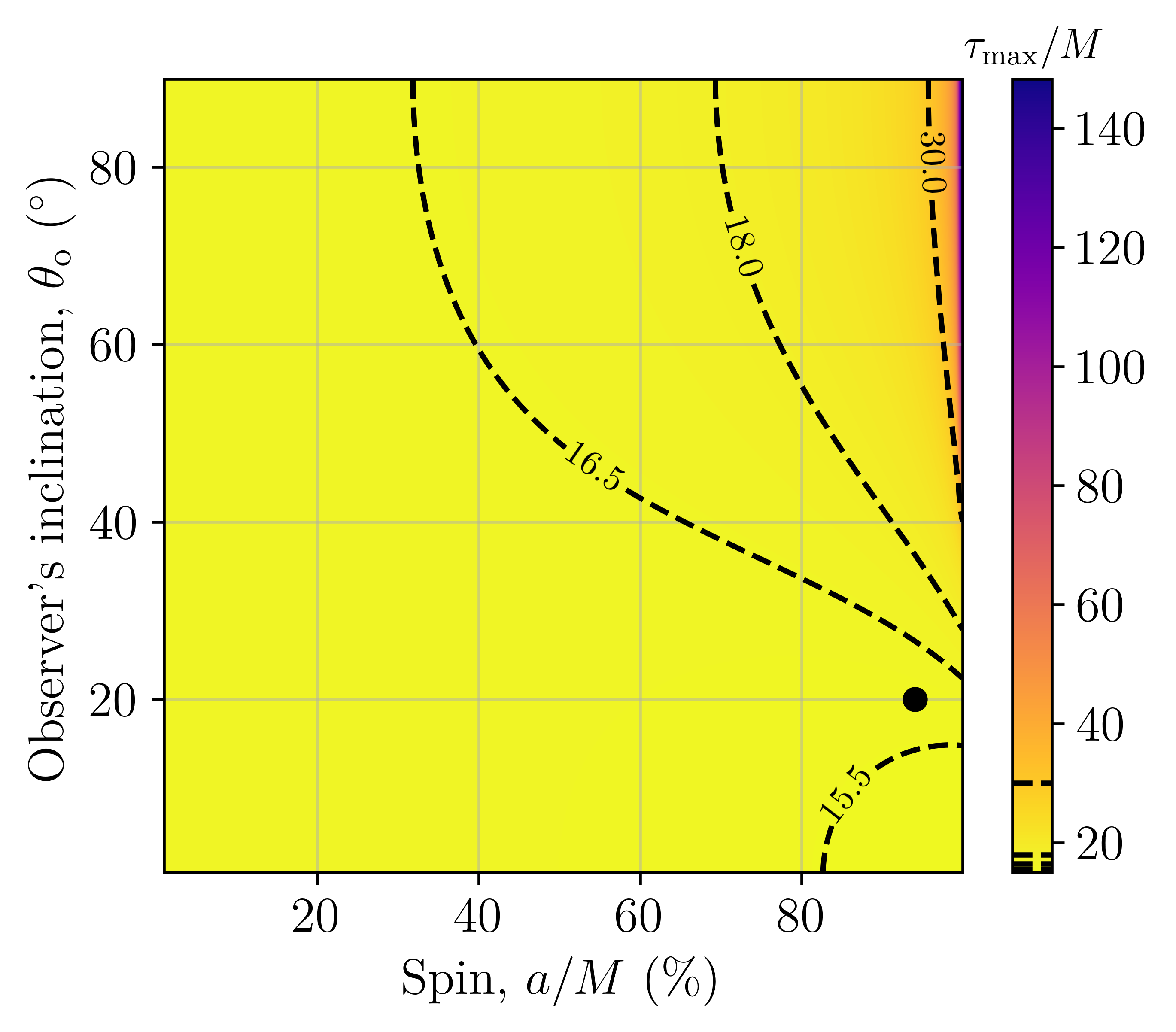}
	\includegraphics[width=0.4\columnwidth]{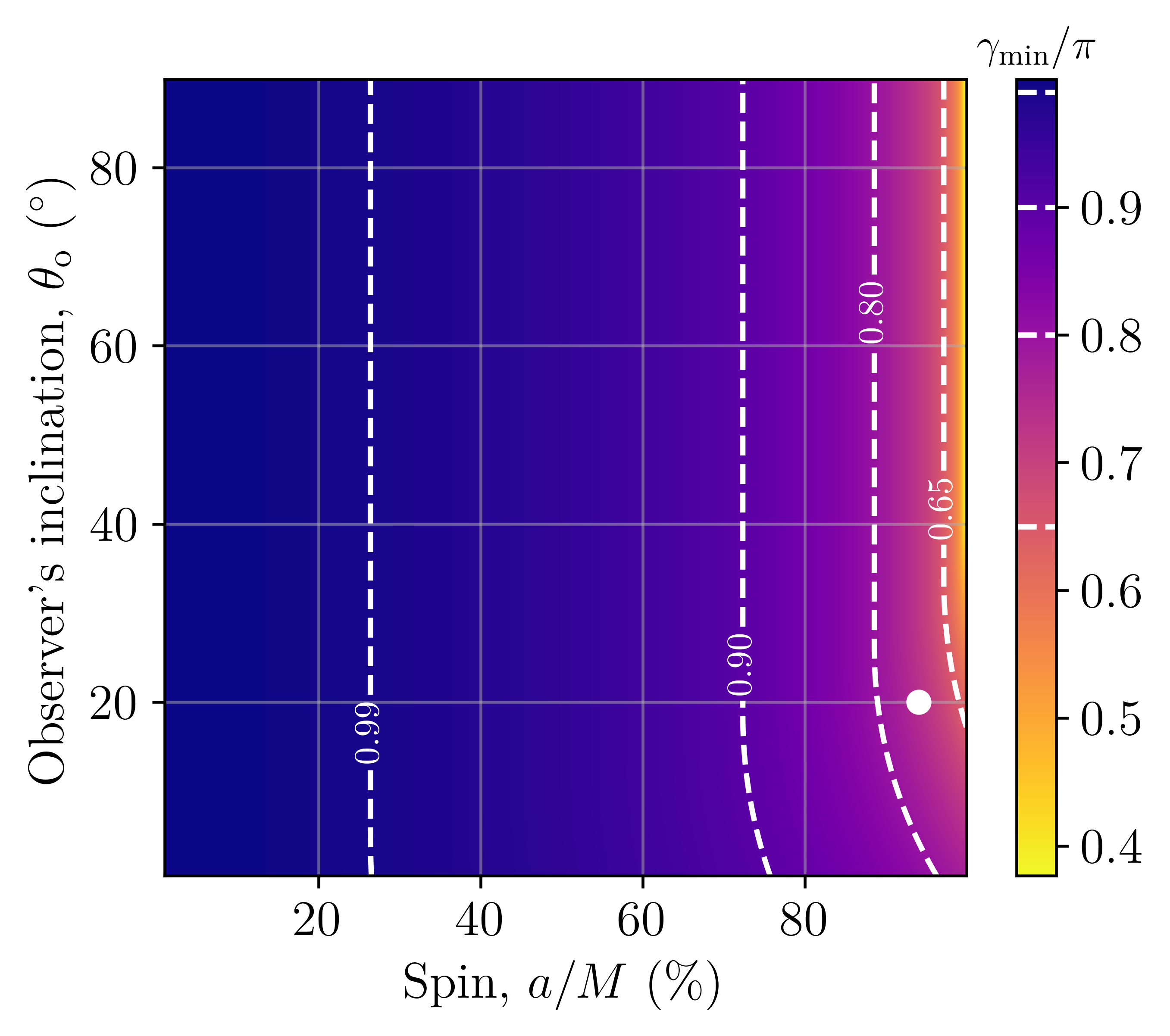}
    \caption{\textbf{Left:} The maximum value of the time delay $\tau$ (in units of $M$) around the critical curve (or equivalently, the maximum time lapse incurred during a half-orbit within the photon shell), as a function of black hole spin and observer inclination.
    The time delay is $\tau\approx16M$ across most of parameter space, and only grows larger for rapidly spinning black holes observed from high inclinations.
    The black dot represents the location of our fiducial model with spin $a_*=94\%$, inclination $\theta_{\rm o}=20^\circ$, and a maximal half-orbital period $\tau_{\rm max}=15.85 M$.
    \textbf{Right:} The minimum value of the Lyapunov exponent $\gamma$ (rescaled by $\pi$) around the critical curve (which controls the demagnification factor $e^{-\gamma}$), as a function of black hole spin and observer inclination.
    This exponent is $\gamma\approx\pi$ across most of parameter space, though it can decrease significantly at high spin.
    The white dot represents our fiducial model with spin $a_*=94\%$, inclination $\theta_{\rm o}=20^\circ$, and a minimum Lyapunov exponent $\gamma_{\rm min}=0.74\pi$.}
    \label{fig:MaxTauMinGamma}
\end{figure}

As discussed in the main text, the presence of a black hole in the midst of an astrophysical source such as a hot radiating plasma is expected to produce light echoes in its observational appearance.\footnote{The light echoes that we study herein are produced by orbiting light that is lensed into the photon ring.
Such echoes are different from those produced by X-ray ``reverberations'' that are reflected from different parts of the accretion disk before traveling directly to the observer (along $n=0$ rays), and for which there is now empirical evidence \cite{Wilkins2021}.
The time delay between such echoes is not controlled by the period of photon orbits, but rather by the distance between the parts of the disk reflecting X-ray emission.}
The time interval between light echoes can depend on the details of the source, as well as on the black hole mass, spin, and inclination.
However, for strongly lensed echoes, the specific value of this time delay is fairly insensitive to these parameters.
For instance, as shown in the left panel of Fig.~\ref{fig:MaxTauMinGamma}, the time lapse $\tau$ incurred per half orbit is around $16M$ for most values of spin and inclination \cite{Fukumura2008,Moriyama2019,GrallaLupsasca2020a}, only exceeding $30M$ when the black hole is near extremality and observed from high inclination.
Assuming that $\tau$ takes a single value is a conservative approach.
In practice, different sources produce a distribution of time delays in low-order images, broadening out the time autocorrelation and thereby obscuring light echo peaks.

The effects of strong lensing on the time autocorrelation of the light curve in our model, Eq.~\eqref{eq:Autocorrelation}, are controlled by the critical parameters $\gamma$ and $\tau$.
In particular, the relativistic contributions are suppressed by factors of $e^{-\gamma}$.
The right panel of Fig.~\ref{fig:MaxTauMinGamma} illustrates that the expected suppression is approximately $e^{-\pi}\sim4\%$ for most spins and inclinations.

\section{Simulated black hole movies}

In this section, we briefly summarize the key ingredients needed to simulate the black hole movies analyzed in the main text.
We first review how to ray trace black hole images with the \texttt{AART} code \cite{CardenasAvendano2022}, and then describe our method for simulating fluctuating astrophysical sources using the \texttt{inoisy} code \cite{Lee2021}.
Full details are provided in Refs.~\cite{Lee2021,CardenasAvendano2022}.

\textit{Ray tracing.}---%
We ray trace black hole images---frames in a black hole \textit{movie}---following the approach presented in Refs.~\cite{GLM2020,Chael2021,Paugnat2022} and implemented in \texttt{AART} \cite{CardenasAvendano2022}.
This method assumes that the source consists of a thin disk of emitters orbiting along circular-equatorial Keplerian geodesics until they reach the radius of the innermost stable circular orbit (ISCO).
Beyond this radius, they plunge into the black hole according to Cunningham's prescription \cite{Cunningham1975}. 

The intensity at a particular Cartesian position $(\alpha,\beta)$ on the image plane of an observer is computed by (analytically) tracing the corresponding light ray backwards into the emission region.
Each time a ray intersects the equatorial disk at some Boyer-Lindquist coordinates $\mathbf{x}_{\rm s}=\pa{r_{\rm s},\phi_{\rm s},t_{\rm s}}$, the observed intensity $I_{\rm o}(\alpha,\beta)$ increases by an amount that is determined by the source emissivity $I_{\rm s}(\mathbf{x}_{\rm s})$.
The total observed intensity at time $t_{\rm o}$ is then given by \cite{CardenasAvendano2022}
\begin{align}
	\label{eq:ImageIntensity}
	I_{\rm o}(t_{\rm o},\alpha,\beta)=\sum_{n=0}^{N(\alpha,\beta)-1}\zeta_ng^3\pa{r_{\rm s}^{(n)},\alpha,\beta}I_{\rm s}\pa{r_{\rm s}^{(n)},\phi_{\rm s}^{(n)},t_{\rm s}^{(n)}},
\end{align}
where $\mathbf{x}_{\rm s}^{(n)}=\mathbf{x}_{\rm s}^{(n)}(\alpha,\beta)$ denotes the equatorial position where the ray intersects the equatorial plane for the $(n+1)^\text{th}$ time along its backward trajectory from image-plane position $(\alpha,\beta)$ at observer time $t_{\rm o}$, up to a total number $N(\alpha,\beta)$ along its maximal extension outside the event horizon.
The redshift factor $g$ is determined by the motion of the emitters, while $\zeta_n$ is a ``fudge'' factor, which we assume to be equal to $1$ for $n=0$, and $1.5$ for $n\geq1$.
We include this factor to account for the effects of the geometrical thickness of the disk \cite{Chael2021}.
In the notation of Eq.~\eqref{eq:ImageIntensity}, the direct image is labeled by $n=0$, while the photon ring is the collective sum of successive contributions with $n\geq1$.

\textit{Stochastic model for the astrophysical source.}---%
We now briefly describe our stochastic model for the source emissivity $I_{\rm s}(\mathbf{x}_{\rm s})$ in Eq.~\eqref{eq:ImageIntensity}.
Following Ref.~\cite{Lee2021}, we numerically solve the stochastic partial differential equation
\begin{align}
    \label{eq:GeneralSPDE}
	\br{1-\nabla\cdot\mathbf{\Lambda}\pa{\mathbf{x}_{\rm s}}\nabla}\mathcal{I}(\mathbf{x}_{\rm s})=\mathcal{N}\br{\det{\mathbf{\Lambda}}}^\frac{1}{4}\mathcal{W}(\mathbf{x}_{\rm s}),
\end{align}
where $\mathbf{x}_{\rm s}=(t_{\rm s},x_{\rm s},y_{\rm s})$ are Cartesian coordinates in the equatorial source (hence the subscript `$s$'), $\mathcal{N}$ is a normalization constant, $\mathcal{W}$ is a standard Gaussian white noise process, and the matrix $\mathbf{\Lambda}$ controls the local covariance of the source.
In general, the resulting Gaussian random field (GRF) $\mathcal{I}\pa{\mathbf{x}_{\rm s}}$ can be inhomogeneous and anisotropic, depending on the nature of the position-dependent anisotropies and correlation lengths introduced through $\mathbf{\Lambda}$, which takes the form
\begin{align}
	\label{eq:inoisyCovariance}
	\mathbf{\Lambda}=\sum_{i=0}^2\lambda_i^2\mathbf{u}_i\mathbf{u}_i^T.
\end{align}
Here, the vector $\mathbf{u}_0$ sets the temporal correlation structure of the flow, with $\lambda_0$ its characteristic correlation time, while $\mathbf{u}_1$ and $\mathbf{u}_2$ set its spatial structure, with $\lambda_1$ and $\lambda_2$ the corresponding (spatial) characteristic correlation lengths. 

Given a realization of the GRF---namely, a numerical solution of Eq.~\eqref{eq:GeneralSPDE}---with periodic boundary conditions in time and Dirichlet boundary conditions in space, we create a realization of the ``standardized'' field $\hat{\mathcal{I}}(\mathbf{x}_{\rm s})$ defined by
\begin{align}
	\label{eq:Normalized}
	\hat{\mathcal{I}}(\mathbf{x})=\frac{\mathcal{I}(\mathbf{x})-\av{\mathcal{I}(\mathbf{x})}}{\sqrt{\av{\mathcal{I}^2(\mathbf{x})}-\av{\mathcal{I}(\mathbf{x})}^2}},
\end{align}
where the angle brackets denoting an average over the source.
Then given a choice of parameter $\sigma$ that controls the amplitude of astrophysical fluctuations, we generate a source emissivity \cite{CardenasAvendano2022}
\begin{align}
	\label{eq:Intensity}
	I_{\rm s}(\mathbf{x}_{\rm s})\equiv J_{\rm{SU}}(r_{\rm s})e^{\sigma\hat{\mathcal{I}}(\mathbf{x}_{\rm s})-\frac{1}{2}\sigma^2},
\end{align}
where the function $J_{\rm{SU}}(r_{\rm s})$ serves as an envelope for the radial profile of the fluctuating source.
We take it to be the Johnson's SU distribution, a function with the conveniently simple analytical form
\begin{align}
	\label{eq:JonhnsonSU}
	J_{\rm{SU}}(r;\mu,\vartheta,\gamma)\equiv\frac{e^{-\frac{1}{2}\br{\gamma+\arcsinh\pa{\frac{r-\mu}{\vartheta}}}^2}}{\sqrt{\pa{r-\mu}^2+\vartheta^2}},
\end{align}
whose three parameters $\mu$, $\vartheta$, and $\gamma$ respectively control the location of the profile's peak, its width, and its asymmetry \cite{Paugnat2022}.
Given that the values of these three parameters primarily control the appearance of the time-averaged image, but not the correlations that we focus on here, we set them to the fiducial values $(\mu,\vartheta,\gamma)=(r_-,1/2,-3/2)$, where $r_{-}=M-\sqrt{M^2-a^2}$ denotes the radius of the inner horizon.
As we will show below, this particular set of parameters produces black hole images that are broadly consistent with the 2017 EHT campaign results \cite{GLM2020}.

\section{Realizations of various Gaussian random fields}

In this section, we describe the computation of light curves corresponding to simulated observations of stochastic sources with different correlation structures.
First, we consider an equatorial disk with correlation lengths that remain constant across the disk, and then turn to the case of position-dependent correlations.
For that latter case, we assume that the characteristic timescale $\lambda_0$ of temporal correlations in the source follows a Keplerian scaling $\lambda_0\propto1/\Omega_{\rm K}$, where $\Omega_{\rm K}\propto r_{\rm s}^{3/2}$ denotes the angular velocity of a circular-equatorial Keplerian orbit at radius $r_{\rm s}$ in the source.

\begin{figure*}
    \centering
    \includegraphics[width=\textwidth]{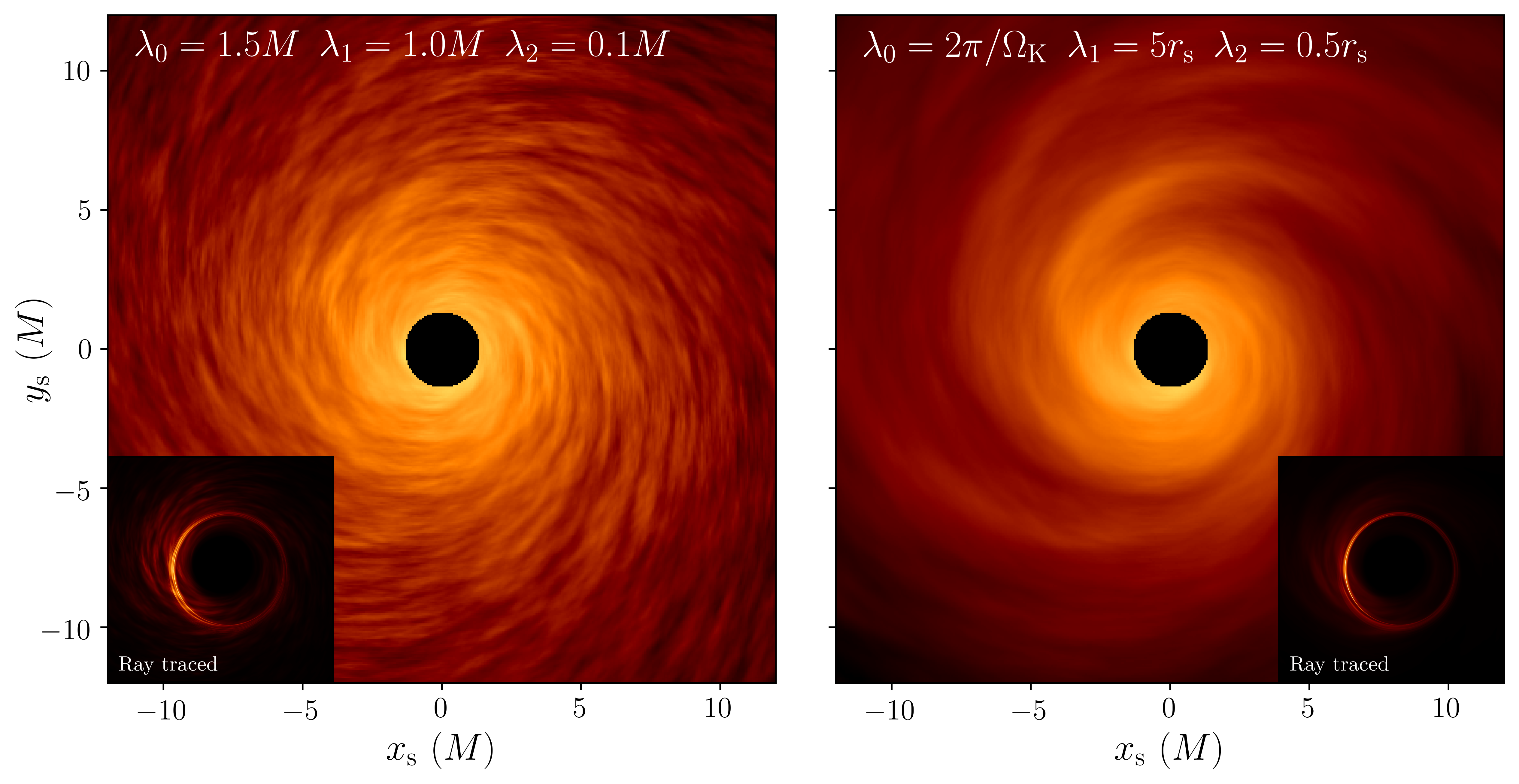}
    \caption{Intensity profiles (in logarithmic scale) of single snapshots from \texttt{inoisy} simulations assuming constant (left) and position-dependent (right) correlation lengths.
    The values for the characteristic scales are shown in the images.
    The inset panels show ray traced images that take into account the time variability of the emission across the image of the disk (i.e., that are ray traced with ``slow light'').
    For both cases, we have used the envelope given in Eq.~\eqref{eq:JonhnsonSU} with $(\mu,\vartheta,\gamma)=(r_-,1/2,-3/2)$, where $r_-=M-\sqrt{M^2-a^2}$ is the inner horizon radius, applied as presented in Eq.~\eqref{eq:Intensity}, with a fluctuation amplitude $\sigma=0.4$.}
    \label{fig:Snapshots}
\end{figure*}

\subsection{Constant-scale correlations}

If the correlation coefficients $\lambda_i$ in the matrix $\mathbf{\Lambda}$ are constant and all equal to the same correlation scale $\lambda$, then Eq.~\eqref{eq:GeneralSPDE} greatly simplifies, and the field it produces is a zero-mean Mat\'ern field with covariance given by \cite{Lindgren2011}
\begin{align}
	\label{eq:MaternCovariance}
	C_\nu(\mathbf{x},\mathbf{y})=\frac{1}{2^{\nu-1}\Gamma(\nu)}\pa{\frac{|\mathbf{x}-\mathbf{y}|}{\lambda}}^\nu K_\nu\pa{\frac{|\mathbf{x}-\mathbf{y}|}{\lambda}},
\end{align}
where $\mathbf{x}$ and $\mathbf{y}$ are three-dimensional position vectors in the equatorial source, $\Gamma(x)$ denotes the Gamma function, and $K_\nu(x)$ is the modified Bessel function of the second kind and order $\nu$.
The order depends on the dimension $d$ of the field and is given by $\nu=2-d/2$. 

As pointed out in Refs.~\cite{Fuglstad2013,Lee2021}, one can numerically construct inhomogeneous and anisotropic GRFs by allowing $\mathbf{\Lambda}$ to vary across the domain.
In particular, following Ref.~\cite{Lee2021}, we take the three-dimensional unit vectors in Eq.~\eqref{eq:inoisyCovariance} to have the following $(t_{\rm s},x_{\rm s},y_{\rm s})$ components:
\begin{align}
    \label{eq:u0}
	\mathbf{u}_0(\mathbf{x}_{\rm s})&=\pa{1,v_x(\mathbf{x}_{\rm s}),v_y(\mathbf{x}_{\rm s})},\\
	\mathbf{u}_1(\mathbf{x}_{\rm s})&=\pa{0,\cos{\theta(\mathbf{x}_{\rm s})},\sin{\theta(\mathbf{x}_{\rm s})}},\\
	\mathbf{u}_2(\mathbf{x}_{\rm s})&=\pa{0,-\sin{\theta(\mathbf{x}_{\rm s})},\cos{\theta(\mathbf{x}_{\rm s})}}.
\end{align}
The resulting flow has a velocity characterized by the vector $\mathbf{v}=\hat{\mathbf{z}}\times\mathbf{x}$.
We set $\theta(\mathbf{x}_{\rm s})=\arctan(y_{\rm s},-x_{\rm s})+\theta_\angle$, where $\theta_\angle=20^\circ$ sets the opening angle of spiral arms in the flow relative to the equatorial circles of constant radius $r_{\rm s}$. 

While the resulting field is still a well-defined GRF arising as a solution to the stochastic partial differential equation given in Eq.~\eqref{eq:GeneralSPDE}, it no longer has a simple analytically known covariance.
Nevertheless, the resulting field can still be well-approximated by an anisotropic Mat\'ern-like covariance function of the form
\begin{align}
	\label{eq:GeneralMaternCovariance}
	C_\nu(\Delta\mathbf{x})=\frac{1}{2^{\nu-1}\Gamma(\nu)}s\pa{\Delta\mathbf{x}}^\nu K_\nu\pa{s\pa{\Delta\mathbf{x}}},
\end{align}
with $\Delta\mathbf{x}=|\mathbf{x}-\mathbf{y}|$ and $s(\Delta\mathbf{x})^2=\Delta\mathbf{x}\cdot\mathbf{\Lambda}^{-1}\Delta\mathbf{x}$.

With the above choices for the correlation structure of the disk, one can generate an inhomogeneous, anisotropic, time-dependent GRF $\mathcal{I}(\mathbf{x}_{\rm s})$, and then produce an accretion-like source, provided that one applies to the GRF an envelope to control the behavior of the observed intensity.
As mentioned in the previous section, we take the source to be given by Eq.~\eqref{eq:Intensity}, choosing an amplitude $\sigma=0.4$ for the fluctuations and using as an envelope the function given in Eq.~\eqref{eq:JonhnsonSU} with $(\mu,\vartheta,\gamma)=(r_-,1/2,-3/2)$, where $r_-=M-\sqrt{M^2-a^2}$.
An example of the resulting field for $\lambda_0=1.5M$, $\lambda_1=1.0M$ and $\lambda_2=0.1M$ is shown in the left panel of Fig.~\ref{fig:Snapshots}. 

These simulations, and all the ones considered in this work, were run on a regular Cartesian grid $(t_{\rm s},x_{\rm s},y_{\rm s})$ of size $4096\times1024\times 1024$.
Specifically, for each of the spatial coordinates $(x_{\rm s},y_{\rm s})$ in the equatorial disk, we uniformly placed $1024$ pixels within the range $[-30,30]M$, resulting in a resolution of about $0.06M$, whereas for the time coordinate $t_{\rm s}$, we placed $4096$ grid points uniformly distributed within the range $[0,5000]M$, resulting in a cadence of $1.22M$.
In Sec.~\ref{app:Convergence} below, we verify that these resolutions are sufficient to correctly compute the autocorrelations, and to ensure that we are resolving all the relevant characteristic scales. 

Despite the intricacies involved in analytically describing the behavior of inhomogeneous and anisotropic fields, if $\lambda_0$ is a constant, then one can still create a simple analytical model for the autocorrelation of the light curve $\mathscr{L}(t)$. 

As shown in Sec.~IIID of Ref.~\cite{CardenasAvendano2022}, one can compute several properties of these GRFs by going to momentum space via the Fourier transform
\begin{align}
	\label{eq:FourierConvention}
	\tilde{\mathcal{I}}(\mathbf{k})=\int\mathcal{I}(\mathbf{x})e^{-i\mathbf{k}\cdot\mathbf{x}}\ed^3\mathbf{x}.
\end{align}
In the Fourier domain, one can write the autocorrelation, given by Eq.~\eqref{eq:LightCurveCorrelationAux}, as
\begin{align}
	\av{\tilde{\mathscr{L}}^*(k_0)\tilde{\mathscr{L}}\pa{k_0^\prime}}=\av{\tilde{\mathcal{I}}^*(\mathbf{k})\tilde{\mathcal{I}}\pa{\mathbf{k}^\prime}},
\end{align}
where the asterisks denote complex conjugation, and we have taken the momentum vector to be $\mathbf{k}=(k_0,0,0)$.
If the field $\mathcal{I}\pa{\mathbf{x}}$ is homogeneous, and we let $\mathbf{x}^\prime=\mathbf{x}+\Delta\mathbf{x}$, then its covariance in Fourier space is given by \cite{CardenasAvendano2022}
\begin{align}
    \av{\tilde{\mathcal{I}}^*(\mathbf{k})\tilde{\mathcal{I}}\pa{\mathbf{k}'}}=2\pi\delta\pa{k_0-k_0'}\tilde{C}_{\nu}\pa{k_0},
\end{align}
where $\tilde{C}_\nu\pa{\mathbf{k}}$ is the momentum-space covariance of the field in Eq.~\eqref{eq:MaternCovariance}.
Combining the previous two equations yields 
\begin{align}
	\av{\tilde{\mathscr{L}}^*(k_0)\tilde{\mathscr{L}}\pa{k_0^\prime}}=2\pi\delta\pa{k_0-k_0'}\tilde{C}_{\nu}\pa{k_0}.
\end{align}
Transforming this expression back to the time domain, one obtains Eq.~\eqref{eq:AutocorrelationHomogenenous} in the main text, after evaluating Eq.~\eqref{eq:MaternCovariance} for only one dimension (the temporal one), with $d=1$ and therefore $\nu=3/2$, and setting $\lambda=\lambda_0$.

\begin{figure}
    \centering
    \includegraphics[width=0.45\columnwidth]{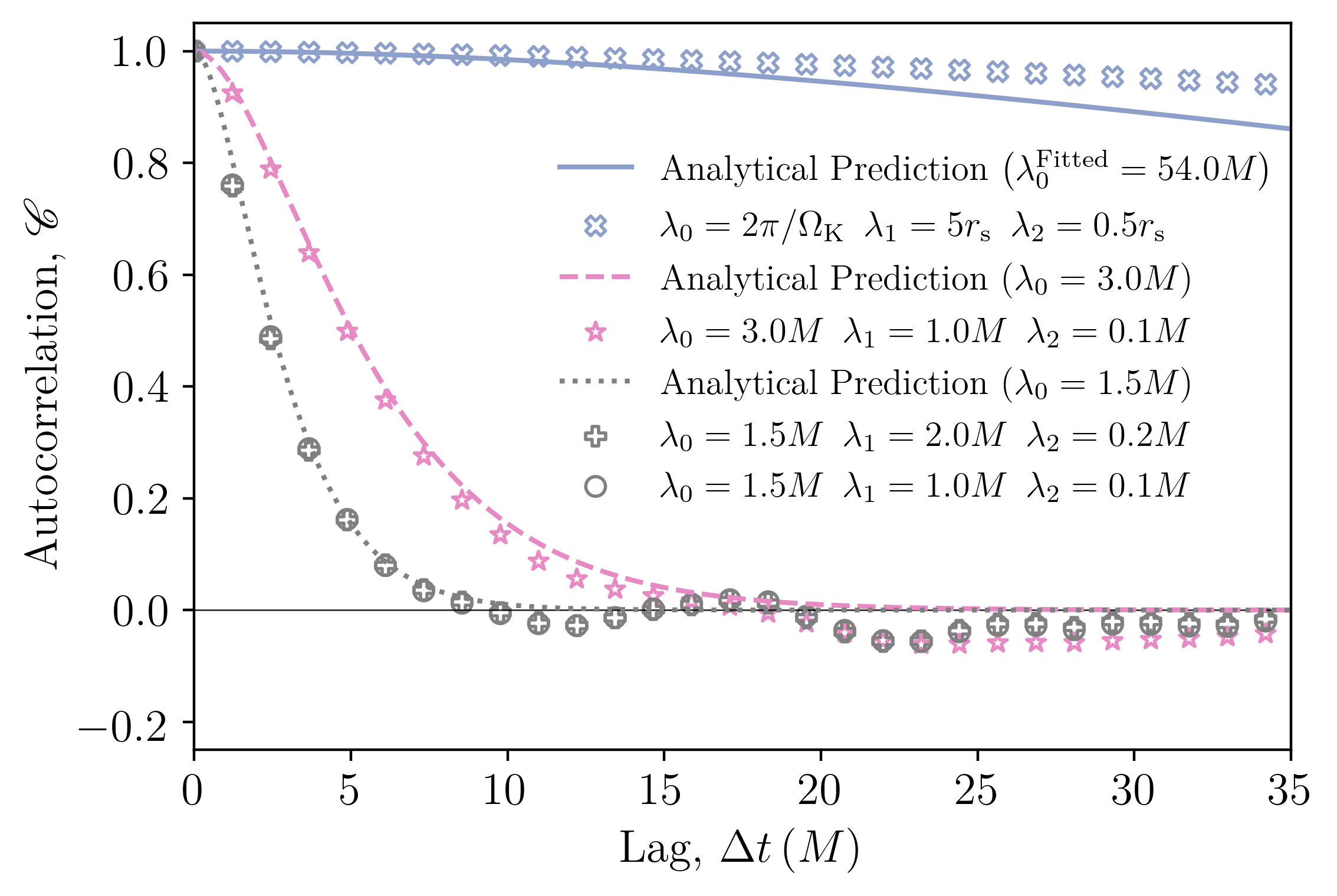}
    \caption{The light curve time autocorrelation resulting from different values of the temporal ($\lambda_0$) and spatial ($\lambda_1$ and $\lambda_2$) correlations of the GRF produced via Eq.~\eqref{eq:Intensity}.
    The lines show the derived analytical prediction, Eq.~\eqref{eq:AutocorrelationHomogenenous}, while the points show the numerical results.
    The blue crosses show the autocorrelation for a Keplerian flow and the line corresponds to a fitted value of $\lambda_0$, which we treat as an effective correlation timescale.
    The spatial correlations do not affect the time autocorrelation of these light curves, as shown by the circles and crosses, which have different spatial correlations, but equal time correlations.}
    \label{fig:LotsOfCorrelations}
\end{figure}

More generally, given a $d$-dimensional Mat\'ern field, it is interesting to note that integrating out $k<d$ of its dimensions results (via essentially the same argument) in a $(d-k)$-dimensional Mat\'ern field.

In Fig.~\ref{fig:LotsOfCorrelations}, we compare our analytical prediction from Eq.~\eqref{eq:AutocorrelationHomogenenous} for the light curve time autocorrelation (shown with lines) against the numerically computed results obtained from GRF realizations solving Eq.~\eqref{eq:GeneralSPDE} (shown with points and crosses).
The grey and pink symbols correspond to simulations with $\lambda_0=1.5M$ and $\lambda_0=3.0M$, respectively.
As the figure shows, despite the (unavoidable) numerical and discretization errors (and of course the fact that one is computing realizations that depend on the underlying ``activation'' white noise $\mathcal{W}$), the prediction from Eq.~\eqref{eq:AutocorrelationHomogenenous} works remarkably well.
In Sec.~\ref{app:Convergence} below, we check that an ensemble of realizations follows the expected statistics and that the numerical errors are under control.
Lastly, the grey circles and crosses in Fig.~\ref{fig:LotsOfCorrelations} correspond to two simulations with spatial correlation length scales, $\lambda_1$ and $\lambda_2$ differing by a factor of two. 
Despite this change, the resulting time autocorrelation is unaffected, as expected since these simulations share the same temporal correlation timescale $\lambda_0$. 

\subsection{Position-dependent correlations}
\label{sec:variableparams}

Although the choice of envelope and constant correlation structure in the previous example is enough to produce an accretion-like emission model, such a model is to simple to reproduce the features that appear in physically motivated models or in observational data.
However, this model becomes capable of qualitatively reproducing several of these features, provided that one allows the characteristic correlation scales $\lambda_i$ to vary across the simulation domain \cite{Lee2021,CardenasAvendano2022}. 

In particular, if we choose the velocity of the flow to follow a Keplerian profile by setting $\lambda_0\propto 1/\Omega_{\rm K}$, and moreover allow the 
correlation lengths to be proportional to
the source radius (so $\lambda_1,\lambda_2\propto r_{\rm s}$), then the resulting ray traced images (shown in the inset in the right panel of Fig.~\ref{fig:Snapshots}) are remarkably similar to state-of-the-art GRMHD snapshots.

As mentioned in the main text, when these parameters are not constant, we cannot derive an analytical expression for the autocorrelation such as the one given in Eq.~\eqref{eq:AutocorrelationHomogenenous}.
We can, however, fit the model in Eq.~\eqref{eq:AutocorrelationHomogenenous} to the numerically computed autocorrelation to obtain an effective timescale $\lambda_0$ of temporal correlations.
For $\lambda_0=2\pi/\Omega_{\rm K}$, $\lambda_1=5r_{\rm s}$, and $\lambda_1=0.5 r_{\rm s}$ (a snapshot of the resulting field is shown in the right panel of Fig.~\ref{fig:Snapshots}), we find an effective value for the correlation timescale of $\lambda_0=88M$, as shown in Fig.~\ref{fig:LotsOfCorrelations} with the blue lines and points.
This effective timescale of the source correlations is expected to be different from the one inferred from observations of the source lensed by the black hole ($\lambda_0^{\rm fitted}=54 M$), whose presence relativistically broadens the autocorrelation.
Although this broadening does not provide a clear signature of the light echo (nor a correlation at the relevant timescale), if one \textit{knew} the time correlation of the underlying emission, then one \textit{could} indirectly infer the presence of the black hole due to this shift; for further discussion of this point, see Sec.~\ref{app:PowerSpectrum} below.

We stress that our model for the autocorrelation of the light curve, Eq.~\eqref{eq:Autocorrelation}, is valid for \textit{any} autocorrelation $\mathscr{C}_0(\Delta t)$ of the underlying emission.
In the main text, we gave a general argument for the behavior of the model when $\tau \gg \lambda_0$ [Eq.~\eqref{eq:Case1}] and when $\tau \ll \lambda_0$ [Eq.~\eqref{eq:Explanation}].
The precise behavior in the intermediate regime $\tau \sim \lambda_0$ depends on the particular functional form of the covariance of the source.
For instance, if $\mathscr{C}_0=\exp\left(-|\Delta t|/\lambda_0\right)$, then the light curve autocorrelation does not display a secondary maximum at $\Delta t\approx\tau$ when $\tau$ is less than $\lambda_0\cosh^{-1}(e^\gamma/2)$.
For $\gamma\approx\pi$, this means that, for this particular choice of autocorrelation, there cannot be a secondary maximum when $\tau\leq3.14\lambda_0$. 

As we argued in the main text, the absence of correlation peaks in such cases cannot be attributed to limitations in the computation of the light curves.
For instance, the sensitivity (or relative change) of $\mathscr{C}$ is exceedingly low for the autocorrelation considered above when $\tau/\lambda_0\ll 1$.
In fact, it is exponentially insensitive, since
\begin{align}
	\left.\frac{d\ln\mathscr{C}}{d\ln \tau}\right|_{\Delta t=\tau}\approx-2\pa{\frac{\tau}{\lambda_0}}^2e^{-\gamma}.
\end{align}

\subsection{Convergence of the simulations}
\label{app:Convergence}

\begin{figure}
    \centering
    \includegraphics[width=0.45\columnwidth]{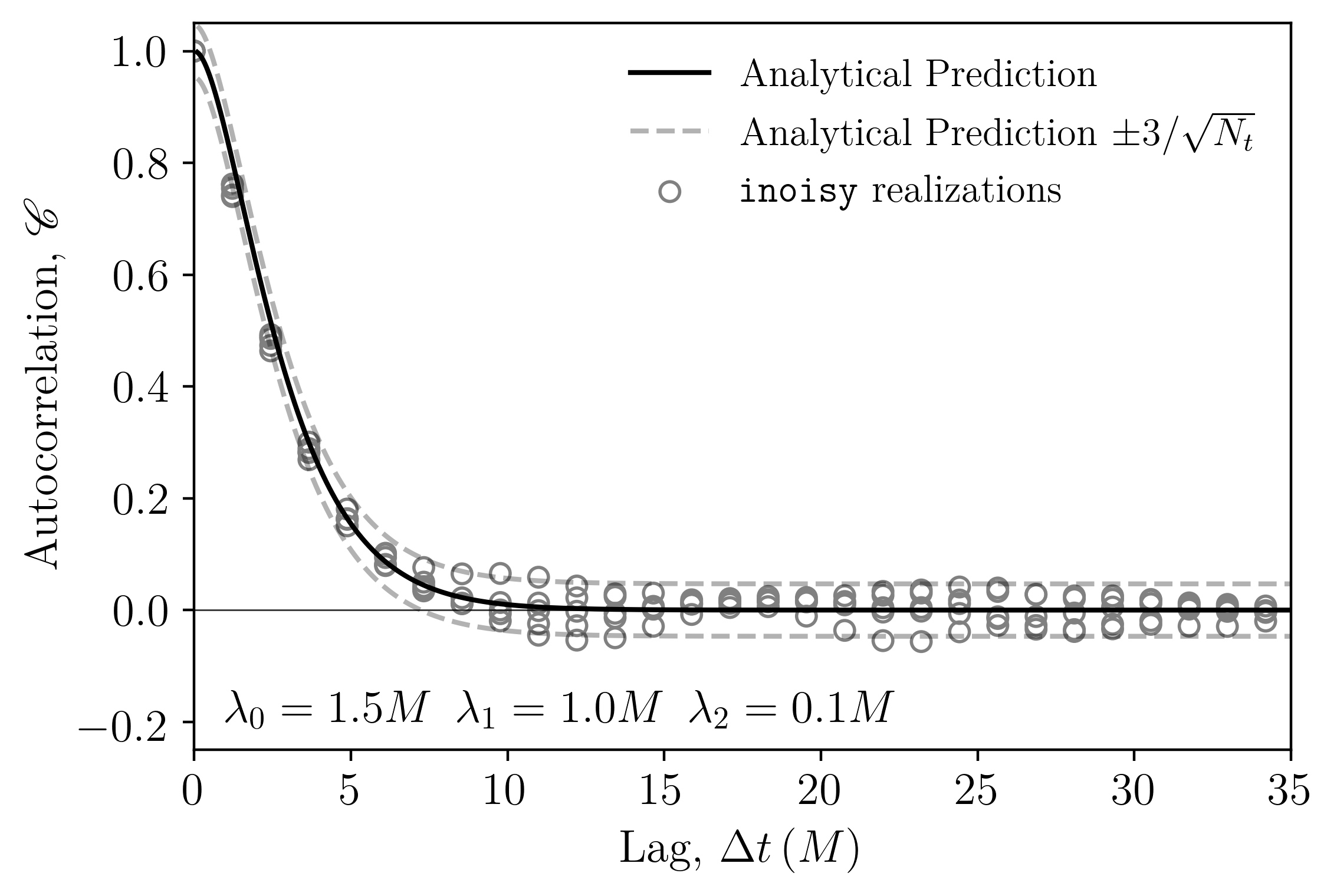}
    \includegraphics[width=0.45\columnwidth]{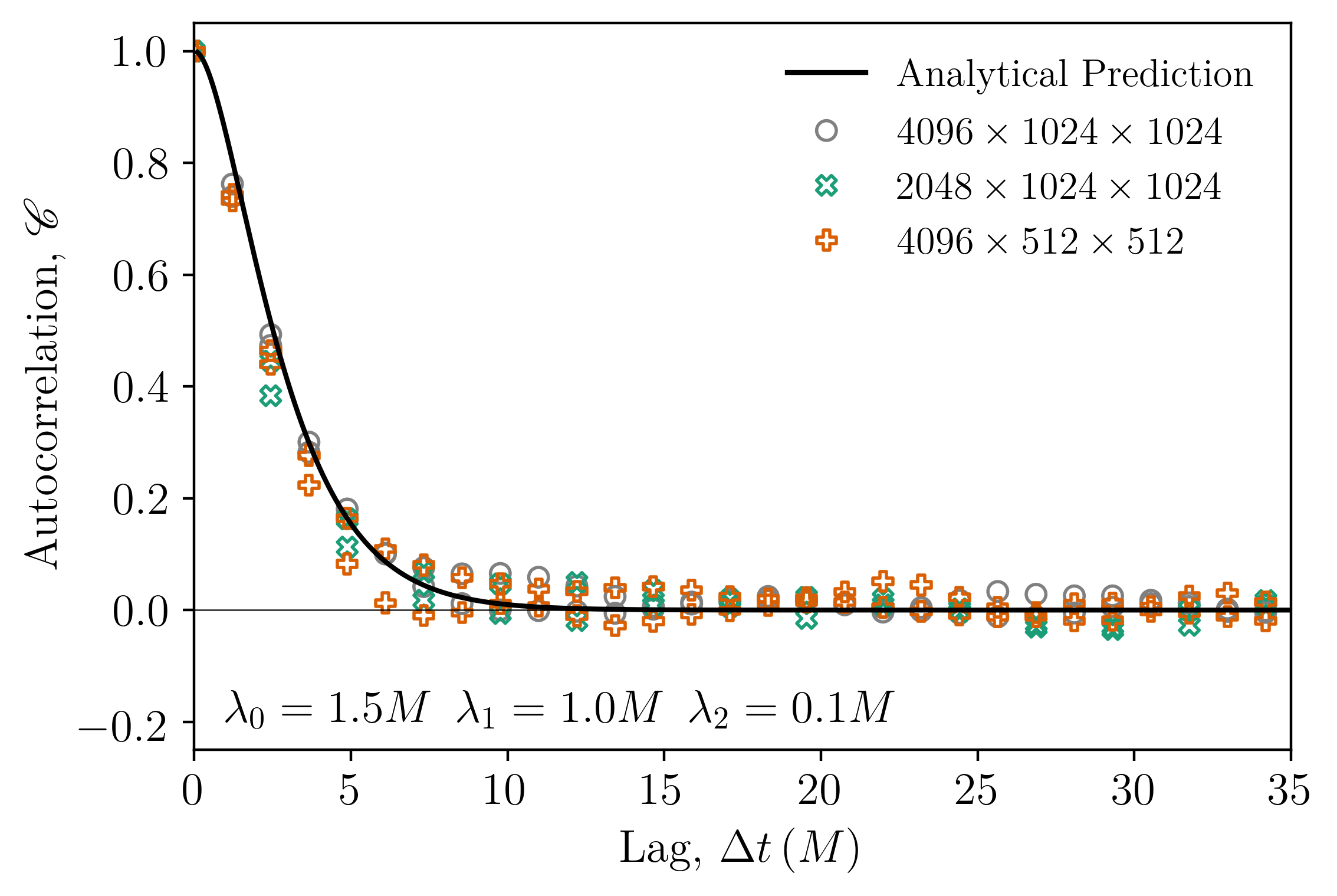}
    \caption{\textbf{Left:} Five different realizations of the GRF obtained from numerically solving Eq.~\eqref{eq:GeneralSPDE} for the same underlying correlation structure and resolution.
    Their variance at large time lags is well approximated by $1/\sqrt{N_t}$, where $N_t$ corresponds to the number of points in the temporal dimension.
    \textbf{Right:} Different temporal and spatial resolutions for a given realization of the underlying emission structure.
    The resolutions are quoted as $N_{t}\times N_{x}\times N_{y}$.
    Once the temporal resolution is higher than the underlying characteristic timescale (i.e., $dt<\lambda_0$), the resulting autocorrelation is computed accurately.}
    \label{fig:Convergence}
\end{figure}

Our underlying emission models are sourced by an ``activation'' process: the white noise $\mathcal{W}$ on the right-hand side of Eq.~\eqref{eq:GeneralSPDE}.
As a result, each realization of the GRF is distinct, yet still adheres to the underlying model covariance.
The impact of this variation on our observable, the time autocorrelation, is shown in the left panel of Fig.~\ref{fig:Convergence}, where we have generated five realizations for the same underlying model and resolution.
As shown with dashed lines, their variance is well approximated by $1/\sqrt{N_t}$, where $N_t$ corresponds to the number of points in the temporal dimension. 

For the light curve, the spatial components of the underlying simulation are subdominant, so one only needs to resolve the temporal scale correctly.
This is achieved whenever the temporal resolution is lower than the underlying characteristic timescale (i.e., $dt<\lambda_0$), as we show in the right panel of Fig.~\ref{fig:Convergence}.
Therein, we can see that once this condition is met, the resulting autocorrelation is unchanged, modulo the inherent stochastic variations just discussed. 

\section{Autocorrelations from observations of Sagittarius~A*}

For the simulations discussed in the main text, we have tuned the parameters of the underlying \texttt{inoisy} simulation to mimic the high-cadence GRMHD-simulated light curve shown in the top panel of Fig.~\ref{fig:All} in the main text.
The resulting simulated light curves do not qualitatively match the observed Sgr\,A$^*$ light curve: this is the so-called ``variability crisis'' \cite{EHT2022a}. Nevertheless, thanks to the flexibility of the \texttt{inoisy} model, we could also tune it to match the Sgr\,A$^*$ light curve instead by decreasing the scale $\sigma$ of fluctuations.
We stress, however, that $\sigma\sim C(0)$ scales out of the autocorrelation $\mathscr{C}(\delta T)=C(\delta T)/C(0)$, as can be seen from the right panel of Fig.~\ref{fig:All} in the main text.

For the unevenly sampled light curve of Sgr\,A$^*$, we computed its time autocorrelation using the locally normalized discrete correlation function (LNDCF), defined as~\cite{Lehar1992,Edelson1988}
\begin{align}
    \textrm{LNDCF}(\Delta t)=\frac{1}{N_{\Delta t}}\sum_{i,j}\frac{\pa{a_i-\bar{a}_{\Delta t}}]\pa{b_j-\bar{b}_{\Delta t}}}{\sqrt{\pa{\sigma_{a\Delta t}^{2}-e_a^{2}}\pa{\sigma_{b\Delta t}^{2}-e_b^{2}}}},
\end{align}
where, following closely the notation of Ref.~\cite{Wielgus2022}, $a_i$ and $b_i$ denote the flux density measurements of the
two compared light curves, and $e_a $ and $e_b$ their estimated errors.
Meanwhile, $\bar{a}_{\Delta t}$, $\bar{b}_{\Delta t}$, $\sigma_{a \Delta t}$ and $\sigma_{b \Delta t}$ are respectively the flux density means and standard deviations, calculated for each lag $\Delta t$ using only the flux density measurements that contribute to the calculation at that particular lag.
Lastly, $N_{\Delta t}$ is the number of data pairs contributing to the lag bin $\Delta t$. 

\begin{figure}
    \centering
    \includegraphics[width=0.43\columnwidth]{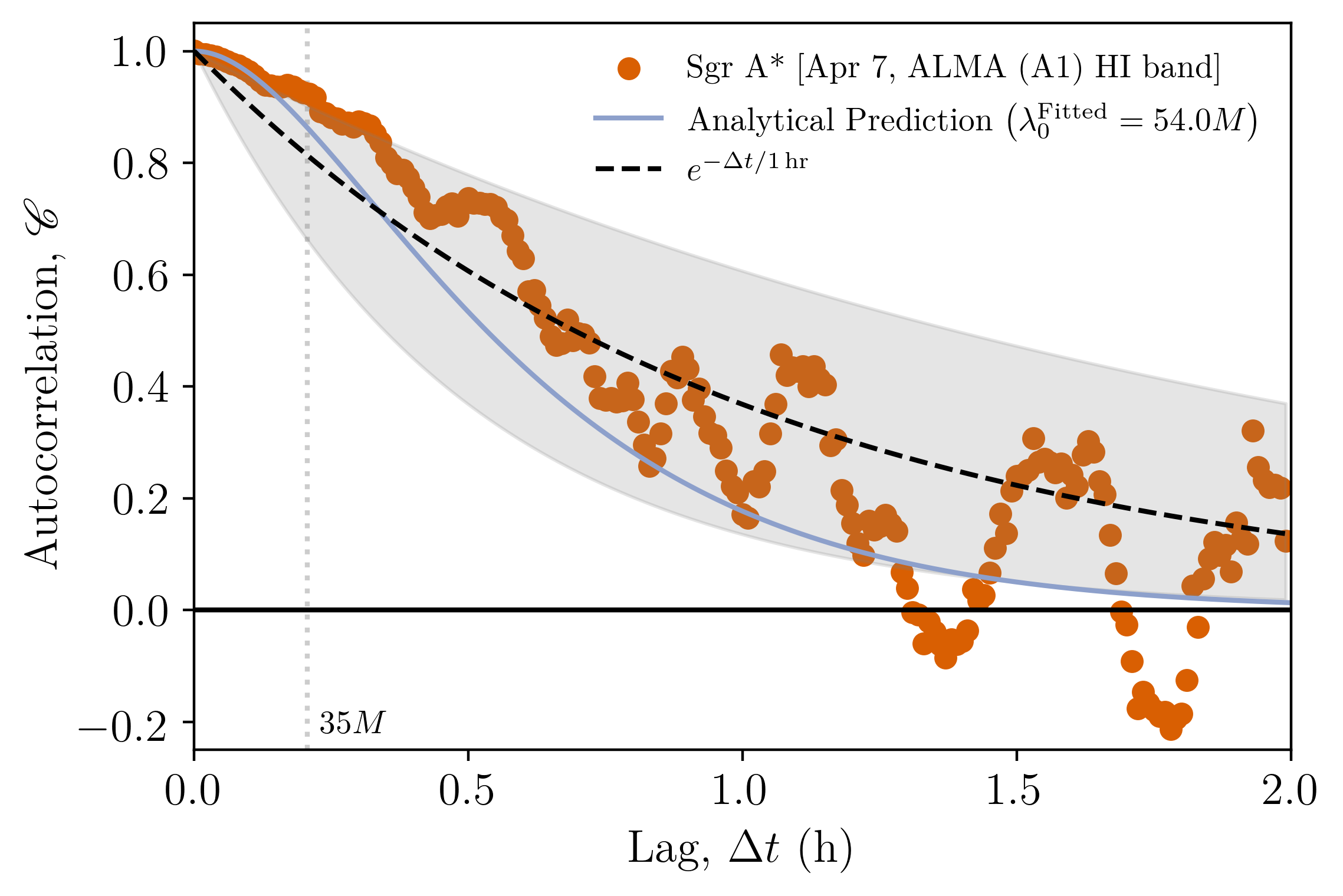}
    \caption{The autocorrelation of the Sgr\,A$^*$ light curve measured on April 7th, 2017.
    As in Ref.~\cite{Wielgus2022}, we plot an exponential decay with a $1$-hour timescale, and the shaded region corresponds to autocorrelation timescales between $0.5$ and $2$ hours.
    This figure can be directly compared with the third panel of Fig.~9 in Ref.~\cite{Wielgus2022}.
    The structure in the resulting autocorrelation can be attributed to the limited and irregular sampling alone \cite{Wielgus2022}.}
    \label{fig:ALMALong}
\end{figure}

As a check of our implementation, in Fig.~\ref{fig:ALMALong}, we reproduce the autocorrelation computed from the ALMA data for April 7th and presented in Ref.~\cite{Wielgus2022}.
As in Ref.~\cite{Wielgus2022}, we have also added a line representing an exponential decay with a $1$-hour timescale, and the shaded region corresponds to autocorrelation timescales between $0.5$ and $2$ hours. 

As discussed in the main text, we have also implemented an alternative way to calculate autocorrelations from the data: using a Gaussian process regression (GPR)~\cite{Rasmussen2006} method to forecast and interpolate the light curve, which allows for an even sampling.
In this way, we can then apply the same procedure we used in the previous sections to compute the simulated autocorrelations.
This approach was pursued in Ref.~\cite{Wielgus2022} to model the measured light curves.  

A GPR uses a covariance function (or kernel) to shape the prior and posterior distributions of the Gaussian process.
For the observed light curve of Sgr\,A$^*$, we implemented two kernels (the Mat\'ern and rational quadratic kernels) and fitted their (hyper)parameters.
As seen in Eq.~\ref{eq:MaternCovariance}, the Mat\'ern covariance (kernel) is characterized by $\nu$ and a length-scale parameter $\lambda$.
By contrast, the rational quadratic kernel serves as a scale mixture of several radial basis function kernels, parameterized by a length-scale $\lambda$ and a scale mixture parameter $\alpha$.
To both models, we have added a constant kernel to scale their magnitude.
The values of the parameters in these kernels are found by maximizing the log-marginal-likelihood; that is, we optimize the parameter when fitting the GPR to the data. In the top panel of Fig.~\ref{fig:All} in the main text, we show the measured light curve, and the resulting two predictions, with their respective error for the gaps. Both of these kernels yield a similar log-likehood ($\sim 10^4$).

\section{Power spectrum of light curve autocorrelations}
\label{app:PowerSpectrum}

Here, we briefly discuss the characteristic signatures that light echoes introduce in the power spectrum (Fourier transform) of the time autocorrelation $\mathcal{C}(\Delta t)$.
Suppose as in Eq.~\eqref{eq:Autocorrelation} that the observed light curve has autocorrelation
\begin{align}
    \label{eq:AutocorrelationBis}
    \mathscr{C}(\Delta t)=\mathscr{C}_0(\Delta t)\br{1-2e^{-\gamma}\mathscr{C}_0(\tau)}+e^{-\gamma}\br{\mathscr{C}_0(\Delta t+\tau)+\mathscr{C}_0(\Delta t-\tau)}+\mathcal{O}\pa{e^{-2\gamma}},
\end{align}
where $\mathscr{C}_0(\Delta t)$ denotes the time autocorrelation of the direct ($n=0$) emission, with Fourier transform
\begin{align}
    \tilde{\mathscr{C}}_0(k)=\int\mathscr{C}_0(\Delta t)e^{-ik\Delta t}\ed\Delta t,\qquad
    \mathscr{C}_0(\Delta t)=\frac{1}{2\pi}\int\tilde{\mathscr{C}}_0(k)e^{+ik\Delta t}\ed k.
\end{align}
Then, the Fourier transform of the full autocorrelation $\mathscr{C}(\Delta t)$ is
\begin{align}
    \tilde{\mathscr{C}}(k)&=\tilde{\mathscr{C}}_0(k)\br{1-2e^{-\gamma}\mathscr{C}_0(\tau)}+e^{-\gamma}\br{e^{+ik\tau}\tilde{\mathscr{C}}_0(k)+e^{-ik\tau}\tilde{\mathscr{C}}_0(k)}+\mathcal{O}\pa{e^{-2\gamma}}.
\end{align}
Thus, we see that the autocorrelation power spectrum in the presence of the black hole, $\tilde{\mathscr{C}}(k)$, differs from the power spectrum in the absence of a black hole, $\tilde{\mathscr{C}}_0(k)$, by a multiplicative term
\begin{align}
    \label{eq:LensedPowerSpectrum}
    \frac{\tilde{\mathscr{C}}(k)}{\tilde{\mathscr{C}}_0(k)}=1+2e^{-\gamma}\br{\cos\pa{\tau k}-\mathscr{C}_0(\tau)}+\mathcal{O}\pa{e^{-2\gamma}}.
\end{align}
This calculation only accounts for the first ($n=1$) light echoes.
We expect the $n^\text{th}$-order light echoes to introduce additional modes oscillating with frequency $n\tau$ but exponentially suppressed by $e^{-n\gamma}$.

These characteristic oscillations in the power spectrum may provide a way to detect the presence of light echoes in black hole light curves even when they do not produce secondary peaks in the time autocorrelation.
In practice, however, we expect a detection of the signature \eqref{eq:LensedPowerSpectrum} to be very challenging for a number of reasons, including:\\
\begin{itemize}
    \item First, as mentioned in Sec.~\ref{app:CriticalParameters}, different sources can produce a distribution of time delays in low-order (small-$n$) images, especially at nonzero inclination, so the oscillation $\cos\pa{\tau k}$ will be smeared over a range of $\tau$.
    \item Second, the intrinsic correlation structure of the source cannot be directly observed, so that one can directly access only $\mathscr{C}(\Delta t)$ or $\tilde{\mathscr{C}}(k)$, but not $\mathscr{C}_0(\Delta t)$ nor $\tilde{\mathscr{C}}_0(k)$.
    In other words, the quantity in Eq.~\eqref{eq:LensedPowerSpectrum} is unobservable.
    \item Third, the source power spectrum $\tilde{\mathscr{C}}_0(k)$ may itself have oscillations with frequencies near $\tau$.
    Intrinsic correlations on the same timescale $\lambda\sim\tau$ as the lensing echoes may very well be present in the accretion flow.
\end{itemize}

To sum up, one would have to know the details of the astrophysical source, and in particular its correlation structure $\tilde{\mathscr{C}}_0(k)$, to claim that an observed oscillation with frequency $\tau$ were produced by lensing rather than the source itself.

\clearpage

\bibliographystyle{apsrev4-1}
\bibliography{NoPeaks.bib}

\begin{thebibliography}{45}%
\makeatletter
\providecommand \@ifxundefined [1]{%
 \@ifx{#1\undefined}
}%
\providecommand \@ifnum [1]{%
 \ifnum #1\expandafter \@firstoftwo
 \else \expandafter \@secondoftwo
 \fi
}%
\providecommand \@ifx [1]{%
 \ifx #1\expandafter \@firstoftwo
 \else \expandafter \@secondoftwo
 \fi
}%
\providecommand \natexlab [1]{#1}%
\providecommand \enquote  [1]{``#1''}%
\providecommand \bibnamefont  [1]{#1}%
\providecommand \bibfnamefont [1]{#1}%
\providecommand \citenamefont [1]{#1}%
\providecommand \href@noop [0]{\@secondoftwo}%
\providecommand \href [0]{\begingroup \@sanitize@url \@href}%
\providecommand \@href[1]{\@@startlink{#1}\@@href}%
\providecommand \@@href[1]{\endgroup#1\@@endlink}%
\providecommand \@sanitize@url [0]{\catcode `\\12\catcode `\$12\catcode
  `\&12\catcode `\#12\catcode `\^12\catcode `\_12\catcode `\%12\relax}%
\providecommand \@@startlink[1]{}%
\providecommand \@@endlink[0]{}%
\providecommand \url  [0]{\begingroup\@sanitize@url \@url }%
\providecommand \@url [1]{\endgroup\@href {#1}{\urlprefix }}%
\providecommand \urlprefix  [0]{URL }%
\providecommand \Eprint [0]{\href }%
\providecommand \doibase [0]{http://dx.doi.org/}%
\providecommand \selectlanguage [0]{\@gobble}%
\providecommand \bibinfo  [0]{\@secondoftwo}%
\providecommand \bibfield  [0]{\@secondoftwo}%
\providecommand \translation [1]{[#1]}%
\providecommand \BibitemOpen [0]{}%
\providecommand \bibitemStop [0]{}%
\providecommand \bibitemNoStop [0]{.\EOS\space}%
\providecommand \EOS [0]{\spacefactor3000\relax}%
\providecommand \BibitemShut  [1]{\csname bibitem#1\endcsname}%
\let\auto@bib@innerbib\@empty
\bibitem [{\citenamefont {{Schmidt}}(1963)}]{Schmidt1963}%
  \BibitemOpen
  \bibfield  {author} {\bibinfo {author} {\bibfnamefont {M.}~\bibnamefont
  {{Schmidt}}},\ }\href {\doibase 10.1038/1971040a0} {\bibfield  {journal}
  {\bibinfo  {journal} {\nat}\ }\textbf {\bibinfo {volume} {197}},\ \bibinfo
  {pages} {1040} (\bibinfo {year} {1963})}\BibitemShut {NoStop}%
\bibitem [{\citenamefont {{Bowyer}}\ \emph {et~al.}(1965)\citenamefont
  {{Bowyer}}, \citenamefont {{Byram}}, \citenamefont {{Chubb}},\ and\
  \citenamefont {{Friedman}}}]{Bowyer1965}%
  \BibitemOpen
  \bibfield  {author} {\bibinfo {author} {\bibfnamefont {S.}~\bibnamefont
  {{Bowyer}}}, \bibinfo {author} {\bibfnamefont {E.~T.}\ \bibnamefont
  {{Byram}}}, \bibinfo {author} {\bibfnamefont {T.~A.}\ \bibnamefont
  {{Chubb}}}, \ and\ \bibinfo {author} {\bibfnamefont {H.}~\bibnamefont
  {{Friedman}}},\ }\href {\doibase 10.1126/science.147.3656.394} {\bibfield
  {journal} {\bibinfo  {journal} {Science}\ }\textbf {\bibinfo {volume}
  {147}},\ \bibinfo {pages} {394} (\bibinfo {year} {1965})}\BibitemShut
  {NoStop}%
\bibitem [{\citenamefont {{Narayan}}\ and\ \citenamefont
  {{Quataert}}(2023)}]{Narayan2023}%
  \BibitemOpen
  \bibfield  {author} {\bibinfo {author} {\bibfnamefont {R.}~\bibnamefont
  {{Narayan}}}\ and\ \bibinfo {author} {\bibfnamefont {E.}~\bibnamefont
  {{Quataert}}},\ }\href {\doibase 10.1038/s41586-023-05768-4} {\bibfield
  {journal} {\bibinfo  {journal} {\nat}\ }\textbf {\bibinfo {volume} {615}},\
  \bibinfo {pages} {597} (\bibinfo {year} {2023})},\ \Eprint
  {http://arxiv.org/abs/2303.13229} {arXiv:2303.13229} \BibitemShut {NoStop}%
\bibitem [{\citenamefont {{Luminet}}(1979)}]{Luminet1979}%
  \BibitemOpen
  \bibfield  {author} {\bibinfo {author} {\bibfnamefont {J.~P.}\ \bibnamefont
  {{Luminet}}},\ }\href@noop {} {\bibfield  {journal} {\bibinfo  {journal}
  {\aap}\ }\textbf {\bibinfo {volume} {75}},\ \bibinfo {pages} {228} (\bibinfo
  {year} {1979})}\BibitemShut {NoStop}%
\bibitem [{\citenamefont {{Gralla}}\ \emph {et~al.}(2019)\citenamefont
  {{Gralla}}, \citenamefont {{Holz}},\ and\ \citenamefont
  {{Wald}}}]{Gralla2019}%
  \BibitemOpen
  \bibfield  {author} {\bibinfo {author} {\bibfnamefont {S.~E.}\ \bibnamefont
  {{Gralla}}}, \bibinfo {author} {\bibfnamefont {D.~E.}\ \bibnamefont
  {{Holz}}}, \ and\ \bibinfo {author} {\bibfnamefont {R.~M.}\ \bibnamefont
  {{Wald}}},\ }\href {\doibase 10.1103/PhysRevD.100.024018} {\bibfield
  {journal} {\bibinfo  {journal} {\prd}\ }\textbf {\bibinfo {volume} {100}},\
  \bibinfo {eid} {024018} (\bibinfo {year} {2019})},\ \Eprint
  {http://arxiv.org/abs/1906.00873} {arXiv:1906.00873} \BibitemShut {NoStop}%
\bibitem [{\citenamefont {{Johnson}}\ \emph {et~al.}(2020)\citenamefont
  {{Johnson}}, \citenamefont {{Lupsasca}} \emph
  {et~al.}}]{JohnsonLupsasca2020}%
  \BibitemOpen
  \bibfield  {author} {\bibinfo {author} {\bibfnamefont {M.~D.}\ \bibnamefont
  {{Johnson}}}, \bibinfo {author} {\bibfnamefont {A.}~\bibnamefont
  {{Lupsasca}}},  \emph {et~al.},\ }\href {\doibase 10.1126/sciadv.aaz1310}
  {\bibfield  {journal} {\bibinfo  {journal} {Science Advances}\ }\textbf
  {\bibinfo {volume} {6}},\ \bibinfo {pages} {eaaz1310} (\bibinfo {year}
  {2020})},\ \Eprint {http://arxiv.org/abs/1907.04329} {arXiv:1907.04329}
  \BibitemShut {NoStop}%
\bibitem [{\citenamefont {{Lupsasca}}\ \emph {et~al.}(2024)\citenamefont
  {{Lupsasca}}, \citenamefont {{Mayerson}}, \citenamefont {{Ripperda}},\ and\
  \citenamefont {{Staelens}}}]{Lupsasca2024}%
  \BibitemOpen
  \bibfield  {author} {\bibinfo {author} {\bibfnamefont {A.}~\bibnamefont
  {{Lupsasca}}}, \bibinfo {author} {\bibfnamefont {D.~R.}\ \bibnamefont
  {{Mayerson}}}, \bibinfo {author} {\bibfnamefont {B.}~\bibnamefont
  {{Ripperda}}}, \ and\ \bibinfo {author} {\bibfnamefont {S.}~\bibnamefont
  {{Staelens}}},\ }in\ \href {\doibase 10.48550/arXiv.2402.01290} {\emph
  {\bibinfo {booktitle} {Recent Progress on Gravity Tests: Challenges and
  Future Perspectives}}},\ \bibinfo {editor} {edited by\ \bibinfo {editor}
  {\bibfnamefont {C.}~\bibnamefont {{Bambi}}}\ and\ \bibinfo {editor}
  {\bibfnamefont {A.}~\bibnamefont {{C{\'a}rdenas-Avenda{\~n}o}}}}\ (\bibinfo
  {publisher} {Springer},\ \bibinfo {year} {2024})\ \Eprint
  {http://arxiv.org/abs/2402.01290} {arXiv:2402.01290} \BibitemShut {NoStop}%
\bibitem [{\citenamefont {{Bardeen}}(1973)}]{Bardeen1973}%
  \BibitemOpen
  \bibfield  {author} {\bibinfo {author} {\bibfnamefont {J.~M.}\ \bibnamefont
  {{Bardeen}}},\ }in\ \href@noop {} {\emph {\bibinfo {booktitle} {Black Holes
  (Les Astres Occlus)}}},\ \bibinfo {editor} {edited by\ \bibinfo {editor}
  {\bibfnamefont {C.}~\bibnamefont {{Dewitt}}}\ and\ \bibinfo {editor}
  {\bibfnamefont {B.~S.}\ \bibnamefont {{Dewitt}}}}\ (\bibinfo  {publisher}
  {Gordon and Breach Science Publishers},\ \bibinfo {year} {1973})\ pp.\
  \bibinfo {pages} {215--239}\BibitemShut {NoStop}%
\bibitem [{\citenamefont {{Gralla}}\ and\ \citenamefont
  {{Lupsasca}}(2020)}]{GrallaLupsasca2020a}%
  \BibitemOpen
  \bibfield  {author} {\bibinfo {author} {\bibfnamefont {S.~E.}\ \bibnamefont
  {{Gralla}}}\ and\ \bibinfo {author} {\bibfnamefont {A.}~\bibnamefont
  {{Lupsasca}}},\ }\href {\doibase 10.1103/PhysRevD.101.044031} {\bibfield
  {journal} {\bibinfo  {journal} {\prd}\ }\textbf {\bibinfo {volume} {101}},\
  \bibinfo {eid} {044031} (\bibinfo {year} {2020})},\ \Eprint
  {http://arxiv.org/abs/1910.12873} {arXiv:1910.12873} \BibitemShut {NoStop}%
\bibitem [{\citenamefont {{Teo}}(2003)}]{Teo2003}%
  \BibitemOpen
  \bibfield  {author} {\bibinfo {author} {\bibfnamefont {E.}~\bibnamefont
  {{Teo}}},\ }\href {\doibase 10.1023/A:1026286607562} {\bibfield  {journal}
  {\bibinfo  {journal} {General Relativity and Gravitation}\ }\textbf {\bibinfo
  {volume} {35}},\ \bibinfo {pages} {1909} (\bibinfo {year}
  {2003})}\BibitemShut {NoStop}%
\bibitem [{\citenamefont {{Hadar}}\ \emph {et~al.}(2021)\citenamefont
  {{Hadar}}, \citenamefont {{Johnson}}, \citenamefont {{Lupsasca}},\ and\
  \citenamefont {{Wong}}}]{Hadar2021}%
  \BibitemOpen
  \bibfield  {author} {\bibinfo {author} {\bibfnamefont {S.}~\bibnamefont
  {{Hadar}}}, \bibinfo {author} {\bibfnamefont {M.~D.}\ \bibnamefont
  {{Johnson}}}, \bibinfo {author} {\bibfnamefont {A.}~\bibnamefont
  {{Lupsasca}}}, \ and\ \bibinfo {author} {\bibfnamefont {G.~N.}\ \bibnamefont
  {{Wong}}},\ }\href {\doibase 10.1103/PhysRevD.103.104038} {\bibfield
  {journal} {\bibinfo  {journal} {\prd}\ }\textbf {\bibinfo {volume} {103}},\
  \bibinfo {eid} {104038} (\bibinfo {year} {2021})},\ \Eprint
  {http://arxiv.org/abs/2010.03683} {arXiv:2010.03683} \BibitemShut {NoStop}%
\bibitem [{\citenamefont {{Akiyama}}\ \emph {et~al.}(2019)\citenamefont
  {{Akiyama}} \emph {et~al.}}]{EHT2019a}%
  \BibitemOpen
  \bibfield  {author} {\bibinfo {author} {\bibfnamefont {K.}~\bibnamefont
  {{Akiyama}}} \emph {et~al.} (\bibinfo {collaboration} {Event Horizon
  Telescope Collaboration}),\ }\href {\doibase 10.3847/2041-8213/ab0ec7}
  {\bibfield  {journal} {\bibinfo  {journal} {\apjl}\ }\textbf {\bibinfo
  {volume} {875}},\ \bibinfo {eid} {L1} (\bibinfo {year} {2019})},\ \Eprint
  {http://arxiv.org/abs/1906.11238} {arXiv:1906.11238} \BibitemShut {NoStop}%
\bibitem [{\citenamefont {{Akiyama}}\ \emph {et~al.}(2024)\citenamefont
  {{Akiyama}} \emph {et~al.}}]{EHT2024}%
  \BibitemOpen
  \bibfield  {author} {\bibinfo {author} {\bibfnamefont {K.}~\bibnamefont
  {{Akiyama}}} \emph {et~al.} (\bibinfo {collaboration} {Event Horizon
  Telescope Collaboration}),\ }\href {\doibase 10.1051/0004-6361/202347932}
  {\bibfield  {journal} {\bibinfo  {journal} {\aap}\ }\textbf {\bibinfo
  {volume} {681}},\ \bibinfo {eid} {A79} (\bibinfo {year} {2024})}\BibitemShut
  {NoStop}%
\bibitem [{\citenamefont {{Akiyama}}\ \emph {et~al.}(2022)\citenamefont
  {{Akiyama}} \emph {et~al.}}]{EHT2022a}%
  \BibitemOpen
  \bibfield  {author} {\bibinfo {author} {\bibfnamefont {K.}~\bibnamefont
  {{Akiyama}}} \emph {et~al.} (\bibinfo {collaboration} {Event Horizon
  Telescope Collaboration}),\ }\href {\doibase 10.3847/2041-8213/ac6674}
  {\bibfield  {journal} {\bibinfo  {journal} {\apjl}\ }\textbf {\bibinfo
  {volume} {930}},\ \bibinfo {eid} {L12} (\bibinfo {year} {2022})}\BibitemShut
  {NoStop}%
\bibitem [{\citenamefont {{Gurvits}}\ \emph {et~al.}(2022)\citenamefont
  {{Gurvits}} \emph {et~al.}}]{Gurvits2022}%
  \BibitemOpen
  \bibfield  {author} {\bibinfo {author} {\bibfnamefont {L.~I.}\ \bibnamefont
  {{Gurvits}}} \emph {et~al.},\ }\href {\doibase
  10.1016/j.actaastro.2022.04.020} {\bibfield  {journal} {\bibinfo  {journal}
  {Acta Astronautica}\ }\textbf {\bibinfo {volume} {196}},\ \bibinfo {pages}
  {314} (\bibinfo {year} {2022})},\ \Eprint {http://arxiv.org/abs/2204.09144}
  {arXiv:2204.09144} \BibitemShut {NoStop}%
\bibitem [{\citenamefont {{Kurczynski}}\ \emph {et~al.}(2022)\citenamefont
  {{Kurczynski}} \emph {et~al.}}]{Kurczynski2022}%
  \BibitemOpen
  \bibfield  {author} {\bibinfo {author} {\bibfnamefont {P.}~\bibnamefont
  {{Kurczynski}}} \emph {et~al.},\ }in\ \href {\doibase 10.1117/12.2630313}
  {\emph {\bibinfo {booktitle} {Space Telescopes and Instrumentation 2022:
  Optical, Infrared, and Millimeter Wave}}},\ \bibinfo {series} {Society of
  Photo-Optical Instrumentation Engineers (SPIE) Conference Series}, Vol.\
  \bibinfo {volume} {12180},\ \bibinfo {editor} {edited by\ \bibinfo {editor}
  {\bibfnamefont {L.~E.}\ \bibnamefont {{Coyle}}}, \bibinfo {editor}
  {\bibfnamefont {S.}~\bibnamefont {{Matsuura}}}, \ and\ \bibinfo {editor}
  {\bibfnamefont {M.~D.}\ \bibnamefont {{Perrin}}}}\ (\bibinfo {year} {2022})\
  p.\ \bibinfo {pages} {121800M}\BibitemShut {NoStop}%
\bibitem [{\citenamefont {{Kudriashov}}\ \emph {et~al.}(2021)\citenamefont
  {{Kudriashov}} \emph {et~al.}}]{Kudriashov2021}%
  \BibitemOpen
  \bibfield  {author} {\bibinfo {author} {\bibfnamefont {V.}~\bibnamefont
  {{Kudriashov}}} \emph {et~al.},\ }\href {\doibase
  10.3724/SP.J.0254-6124.2021.0202} {\bibfield  {journal} {\bibinfo  {journal}
  {Chinese Journal of Space Science}\ }\textbf {\bibinfo {volume} {41}},\
  \bibinfo {pages} {211} (\bibinfo {year} {2021})},\ \Eprint
  {http://arxiv.org/abs/2105.06882} {arXiv:2105.06882} \BibitemShut {NoStop}%
\bibitem [{\citenamefont {{Ulrich}}\ \emph {et~al.}(1997)\citenamefont
  {{Ulrich}}, \citenamefont {{Maraschi}},\ and\ \citenamefont
  {{Urry}}}]{Ulrich1997}%
  \BibitemOpen
  \bibfield  {author} {\bibinfo {author} {\bibfnamefont {M.-H.}\ \bibnamefont
  {{Ulrich}}}, \bibinfo {author} {\bibfnamefont {L.}~\bibnamefont
  {{Maraschi}}}, \ and\ \bibinfo {author} {\bibfnamefont {C.~M.}\ \bibnamefont
  {{Urry}}},\ }\href {\doibase 10.1146/annurev.astro.35.1.445} {\bibfield
  {journal} {\bibinfo  {journal} {\araa}\ }\textbf {\bibinfo {volume} {35}},\
  \bibinfo {pages} {445} (\bibinfo {year} {1997})}\BibitemShut {NoStop}%
\bibitem [{\citenamefont {{Remillard}}\ and\ \citenamefont
  {{McClintock}}(2006)}]{Remillard2006}%
  \BibitemOpen
  \bibfield  {author} {\bibinfo {author} {\bibfnamefont {R.~A.}\ \bibnamefont
  {{Remillard}}}\ and\ \bibinfo {author} {\bibfnamefont {J.~E.}\ \bibnamefont
  {{McClintock}}},\ }\href {\doibase 10.1146/annurev.astro.44.051905.092532}
  {\bibfield  {journal} {\bibinfo  {journal} {\araa}\ }\textbf {\bibinfo
  {volume} {44}},\ \bibinfo {pages} {49} (\bibinfo {year} {2006})},\ \Eprint
  {http://arxiv.org/abs/astro-ph/0606352} {arXiv:astro-ph/0606352} \BibitemShut
  {NoStop}%
\bibitem [{\citenamefont {{Berger}}(2014)}]{Berger2014}%
  \BibitemOpen
  \bibfield  {author} {\bibinfo {author} {\bibfnamefont {E.}~\bibnamefont
  {{Berger}}},\ }\href {\doibase 10.1146/annurev-astro-081913-035926}
  {\bibfield  {journal} {\bibinfo  {journal} {\araa}\ }\textbf {\bibinfo
  {volume} {52}},\ \bibinfo {pages} {43} (\bibinfo {year} {2014})},\ \Eprint
  {http://arxiv.org/abs/1311.2603} {arXiv:1311.2603} \BibitemShut {NoStop}%
\bibitem [{\citenamefont {{Fukumura}}\ and\ \citenamefont
  {{Kazanas}}(2008)}]{Fukumura2008}%
  \BibitemOpen
  \bibfield  {author} {\bibinfo {author} {\bibfnamefont {K.}~\bibnamefont
  {{Fukumura}}}\ and\ \bibinfo {author} {\bibfnamefont {D.}~\bibnamefont
  {{Kazanas}}},\ }\href {\doibase 10.1086/587159} {\bibfield  {journal}
  {\bibinfo  {journal} {\apj}\ }\textbf {\bibinfo {volume} {679}},\ \bibinfo
  {pages} {1413} (\bibinfo {year} {2008})},\ \Eprint
  {http://arxiv.org/abs/0712.1084} {arXiv:0712.1084} \BibitemShut {NoStop}%
\bibitem [{\citenamefont {{Moriyama}}\ \emph {et~al.}(2019)\citenamefont
  {{Moriyama}}, \citenamefont {{Mineshige}}, \citenamefont {{Honma}},\ and\
  \citenamefont {{Akiyama}}}]{Moriyama2019}%
  \BibitemOpen
  \bibfield  {author} {\bibinfo {author} {\bibfnamefont {K.}~\bibnamefont
  {{Moriyama}}}, \bibinfo {author} {\bibfnamefont {S.}~\bibnamefont
  {{Mineshige}}}, \bibinfo {author} {\bibfnamefont {M.}~\bibnamefont
  {{Honma}}}, \ and\ \bibinfo {author} {\bibfnamefont {K.}~\bibnamefont
  {{Akiyama}}},\ }\href {\doibase 10.3847/1538-4357/ab505b} {\bibfield
  {journal} {\bibinfo  {journal} {\apj}\ }\textbf {\bibinfo {volume} {887}},\
  \bibinfo {eid} {227} (\bibinfo {year} {2019})},\ \Eprint
  {http://arxiv.org/abs/1910.10713} {arXiv:1910.10713} \BibitemShut {NoStop}%
\bibitem [{\citenamefont {{Chesler}}\ \emph {et~al.}(2021)\citenamefont
  {{Chesler}}, \citenamefont {{Blackburn}}, \citenamefont {{Doeleman}},
  \citenamefont {{Johnson}}, \citenamefont {{Moran}}, \citenamefont
  {{Narayan}},\ and\ \citenamefont {{Wielgus}}}]{Chesler2020}%
  \BibitemOpen
  \bibfield  {author} {\bibinfo {author} {\bibfnamefont {P.~M.}\ \bibnamefont
  {{Chesler}}}, \bibinfo {author} {\bibfnamefont {L.}~\bibnamefont
  {{Blackburn}}}, \bibinfo {author} {\bibfnamefont {S.~S.}\ \bibnamefont
  {{Doeleman}}}, \bibinfo {author} {\bibfnamefont {M.~D.}\ \bibnamefont
  {{Johnson}}}, \bibinfo {author} {\bibfnamefont {J.~M.}\ \bibnamefont
  {{Moran}}}, \bibinfo {author} {\bibfnamefont {R.}~\bibnamefont {{Narayan}}},
  \ and\ \bibinfo {author} {\bibfnamefont {M.}~\bibnamefont {{Wielgus}}},\
  }\href {\doibase 10.1088/1361-6382/abeae4} {\bibfield  {journal} {\bibinfo
  {journal} {Classical and Quantum Gravity}\ }\textbf {\bibinfo {volume}
  {38}},\ \bibinfo {eid} {125006} (\bibinfo {year} {2021})},\ \Eprint
  {http://arxiv.org/abs/2012.11778} {arXiv:2012.11778} \BibitemShut {NoStop}%
\bibitem [{\citenamefont {{Wong}}(2021)}]{Wong2021}%
  \BibitemOpen
  \bibfield  {author} {\bibinfo {author} {\bibfnamefont {G.~N.}\ \bibnamefont
  {{Wong}}},\ }\href {\doibase 10.3847/1538-4357/abdd2d} {\bibfield  {journal}
  {\bibinfo  {journal} {\apj}\ }\textbf {\bibinfo {volume} {909}},\ \bibinfo
  {eid} {217} (\bibinfo {year} {2021})},\ \Eprint
  {http://arxiv.org/abs/2009.06641} {arXiv:2009.06641} \BibitemShut {NoStop}%
\bibitem [{\citenamefont {{Hadar}}\ \emph {et~al.}(2023)\citenamefont
  {{Hadar}}, \citenamefont {{Harikesh}},\ and\ \citenamefont
  {{Chelouche}}}]{Hadar2023}%
  \BibitemOpen
  \bibfield  {author} {\bibinfo {author} {\bibfnamefont {S.}~\bibnamefont
  {{Hadar}}}, \bibinfo {author} {\bibfnamefont {S.}~\bibnamefont {{Harikesh}}},
  \ and\ \bibinfo {author} {\bibfnamefont {D.}~\bibnamefont {{Chelouche}}},\
  }\href {\doibase 10.1103/PhysRevD.107.124057} {\bibfield  {journal} {\bibinfo
   {journal} {\prd}\ }\textbf {\bibinfo {volume} {107}},\ \bibinfo {eid}
  {124057} (\bibinfo {year} {2023})},\ \Eprint
  {http://arxiv.org/abs/2305.11247} {arXiv:2305.11247} \BibitemShut {NoStop}%
\bibitem [{\citenamefont {{Dexter}}\ \emph {et~al.}(2014)\citenamefont
  {{Dexter}}, \citenamefont {{Kelly}}, \citenamefont {{Bower}}, \citenamefont
  {{Marrone}}, \citenamefont {{Stone}},\ and\ \citenamefont
  {{Plambeck}}}]{Dexter2014}%
  \BibitemOpen
  \bibfield  {author} {\bibinfo {author} {\bibfnamefont {J.}~\bibnamefont
  {{Dexter}}}, \bibinfo {author} {\bibfnamefont {B.}~\bibnamefont {{Kelly}}},
  \bibinfo {author} {\bibfnamefont {G.~C.}\ \bibnamefont {{Bower}}}, \bibinfo
  {author} {\bibfnamefont {D.~P.}\ \bibnamefont {{Marrone}}}, \bibinfo {author}
  {\bibfnamefont {J.}~\bibnamefont {{Stone}}}, \ and\ \bibinfo {author}
  {\bibfnamefont {R.}~\bibnamefont {{Plambeck}}},\ }\href {\doibase
  10.1093/mnras/stu1039} {\bibfield  {journal} {\bibinfo  {journal} {\mnras}\
  }\textbf {\bibinfo {volume} {442}},\ \bibinfo {pages} {2797} (\bibinfo {year}
  {2014})},\ \Eprint {http://arxiv.org/abs/1308.5968} {arXiv:1308.5968}
  \BibitemShut {NoStop}%
\bibitem [{\citenamefont {{Wielgus}}\ \emph {et~al.}(2022)\citenamefont
  {{Wielgus}} \emph {et~al.}}]{Wielgus2022}%
  \BibitemOpen
  \bibfield  {author} {\bibinfo {author} {\bibfnamefont {M.}~\bibnamefont
  {{Wielgus}}} \emph {et~al.} (\bibinfo {collaboration} {Event Horizon
  Telescope Collaboration}),\ }\href {\doibase 10.3847/2041-8213/ac6428}
  {\bibfield  {journal} {\bibinfo  {journal} {\apjl}\ }\textbf {\bibinfo
  {volume} {930}},\ \bibinfo {eid} {L19} (\bibinfo {year} {2022})},\ \Eprint
  {http://arxiv.org/abs/2207.06829} {arXiv:2207.06829} \BibitemShut {NoStop}%
\bibitem [{SM()}]{SM}%
  \BibitemOpen
  \href@noop {} {}\bibinfo {note} {See Supplemental Material, which includes
  Refs.~\cite{Wilkins2021, GLM2020,Chael2021,Paugnat2022,
  Cunningham1975,Lindgren2011,Fuglstad2013}, at LINK, for further details on
  the variation of the photon ring critical parameters with black hole spin and
  inclination; our simulation schemes of black hole movies and Gaussian random
  fields; the autocorrelation from observations of Sagittarius A*; and the
  power spectrum of light curve autocorrelations.}\BibitemShut {Stop}%
\bibitem [{\citenamefont {{Porth}}\ \emph {et~al.}(2021)\citenamefont
  {{Porth}}, \citenamefont {{Mizuno}}, \citenamefont {{Younsi}},\ and\
  \citenamefont {{Fromm}}}]{Porth2021}%
  \BibitemOpen
  \bibfield  {author} {\bibinfo {author} {\bibfnamefont {O.}~\bibnamefont
  {{Porth}}}, \bibinfo {author} {\bibfnamefont {Y.}~\bibnamefont {{Mizuno}}},
  \bibinfo {author} {\bibfnamefont {Z.}~\bibnamefont {{Younsi}}}, \ and\
  \bibinfo {author} {\bibfnamefont {C.~M.}\ \bibnamefont {{Fromm}}},\ }\href
  {\doibase 10.1093/mnras/stab163} {\bibfield  {journal} {\bibinfo  {journal}
  {\mnras}\ }\textbf {\bibinfo {volume} {502}},\ \bibinfo {pages} {2023}
  (\bibinfo {year} {2021})},\ \Eprint {http://arxiv.org/abs/2006.03658}
  {arXiv:2006.03658} \BibitemShut {NoStop}%
\bibitem [{\citenamefont {{Conroy}}\ \emph {et~al.}(2023)\citenamefont
  {{Conroy}}, \citenamefont {{Baub{\"o}ck}}, \citenamefont {{Dhruv}},
  \citenamefont {{Lee}}, \citenamefont {{Broderick}}, \citenamefont {{Chan}},
  \citenamefont {{Georgiev}}, \citenamefont {{Joshi}}, \citenamefont
  {{Prather}},\ and\ \citenamefont {{Gammie}}}]{Conroy2023}%
  \BibitemOpen
  \bibfield  {author} {\bibinfo {author} {\bibfnamefont {N.~S.}\ \bibnamefont
  {{Conroy}}}, \bibinfo {author} {\bibfnamefont {M.}~\bibnamefont
  {{Baub{\"o}ck}}}, \bibinfo {author} {\bibfnamefont {V.}~\bibnamefont
  {{Dhruv}}}, \bibinfo {author} {\bibfnamefont {D.}~\bibnamefont {{Lee}}},
  \bibinfo {author} {\bibfnamefont {A.~E.}\ \bibnamefont {{Broderick}}},
  \bibinfo {author} {\bibfnamefont {C.-k.}\ \bibnamefont {{Chan}}}, \bibinfo
  {author} {\bibfnamefont {B.}~\bibnamefont {{Georgiev}}}, \bibinfo {author}
  {\bibfnamefont {A.~V.}\ \bibnamefont {{Joshi}}}, \bibinfo {author}
  {\bibfnamefont {B.}~\bibnamefont {{Prather}}}, \ and\ \bibinfo {author}
  {\bibfnamefont {C.~F.}\ \bibnamefont {{Gammie}}},\ }\href {\doibase
  10.3847/1538-4357/acd2c8} {\bibfield  {journal} {\bibinfo  {journal} {\apj}\
  }\textbf {\bibinfo {volume} {951}},\ \bibinfo {eid} {46} (\bibinfo {year}
  {2023})},\ \Eprint {http://arxiv.org/abs/2304.03826} {arXiv:2304.03826}
  \BibitemShut {NoStop}%
\bibitem [{\citenamefont {{Gravity Collaboration}}(2024)}]{GRAVITY2024}%
  \BibitemOpen
  \bibfield  {author} {\bibinfo {author} {\bibnamefont {{Gravity
  Collaboration}}},\ }\href {\doibase 10.1051/0004-6361/202346926} {\bibfield
  {journal} {\bibinfo  {journal} {\aap}\ }\textbf {\bibinfo {volume} {684}},\
  \bibinfo {eid} {A200} (\bibinfo {year} {2024})},\ \Eprint
  {http://arxiv.org/abs/2401.17764} {arXiv:2401.17764} \BibitemShut {NoStop}%
\bibitem [{\citenamefont {{Lee}}\ and\ \citenamefont
  {{Gammie}}(2021)}]{Lee2021}%
  \BibitemOpen
  \bibfield  {author} {\bibinfo {author} {\bibfnamefont {D.}~\bibnamefont
  {{Lee}}}\ and\ \bibinfo {author} {\bibfnamefont {C.~F.}\ \bibnamefont
  {{Gammie}}},\ }\href {\doibase 10.3847/1538-4357/abc8f3} {\bibfield
  {journal} {\bibinfo  {journal} {\apj}\ }\textbf {\bibinfo {volume} {906}},\
  \bibinfo {eid} {39} (\bibinfo {year} {2021})},\ \Eprint
  {http://arxiv.org/abs/2011.07151} {arXiv:2011.07151} \BibitemShut {NoStop}%
\bibitem [{\citenamefont {{C{\'a}rdenas-Avenda{\~n}o}}\ \emph
  {et~al.}(2023)\citenamefont {{C{\'a}rdenas-Avenda{\~n}o}}, \citenamefont
  {{Lupsasca}},\ and\ \citenamefont {{Zhu}}}]{CardenasAvendano2022}%
  \BibitemOpen
  \bibfield  {author} {\bibinfo {author} {\bibfnamefont {A.}~\bibnamefont
  {{C{\'a}rdenas-Avenda{\~n}o}}}, \bibinfo {author} {\bibfnamefont
  {A.}~\bibnamefont {{Lupsasca}}}, \ and\ \bibinfo {author} {\bibfnamefont
  {H.}~\bibnamefont {{Zhu}}},\ }\href {\doibase 10.1103/PhysRevD.107.043030}
  {\bibfield  {journal} {\bibinfo  {journal} {\prd}\ }\textbf {\bibinfo
  {volume} {107}},\ \bibinfo {eid} {043030} (\bibinfo {year} {2023})},\ \Eprint
  {http://arxiv.org/abs/2211.07469} {arXiv:2211.07469} \BibitemShut {NoStop}%
\bibitem [{\citenamefont {{Wong}}\ \emph {et~al.}(2024)\citenamefont {{Wong}},
  \citenamefont {{Medeiros}}, \citenamefont {{C\'ardenas-Avenda\~no}},\ and\
  \citenamefont {{Stone}}}]{Wong2024}%
  \BibitemOpen
  \bibfield  {author} {\bibinfo {author} {\bibfnamefont {G.~N.}\ \bibnamefont
  {{Wong}}}, \bibinfo {author} {\bibfnamefont {L.}~\bibnamefont {{Medeiros}}},
  \bibinfo {author} {\bibfnamefont {A.}~\bibnamefont
  {{C\'ardenas-Avenda\~no}}}, \ and\ \bibinfo {author} {\bibfnamefont {J.~M.}\
  \bibnamefont {{Stone}}},\ }\href@noop {} {\enquote {\bibinfo {title}
  {{Measuring Black Hole Light Echoes with Very Long Baseline
  Interferometry}},}\ } (\bibinfo {year} {2024}),\ \bibinfo {note} {to
  appear}\BibitemShut {NoStop}%
\bibitem [{\citenamefont {{Wong}}\ \emph {et~al.}(2022)\citenamefont {{Wong}},
  \citenamefont {{Prather}}, \citenamefont {{Dhruv}} \emph
  {et~al.}}]{Wong2022}%
  \BibitemOpen
  \bibfield  {author} {\bibinfo {author} {\bibfnamefont {G.~N.}\ \bibnamefont
  {{Wong}}}, \bibinfo {author} {\bibfnamefont {B.~S.}\ \bibnamefont
  {{Prather}}}, \bibinfo {author} {\bibfnamefont {V.}~\bibnamefont {{Dhruv}}},
  \emph {et~al.},\ }\href {\doibase 10.3847/1538-4365/ac582e} {\bibfield
  {journal} {\bibinfo  {journal} {\apjs}\ }\textbf {\bibinfo {volume} {259}},\
  \bibinfo {eid} {64} (\bibinfo {year} {2022})},\ \Eprint
  {http://arxiv.org/abs/2202.11721} {arXiv:2202.11721} \BibitemShut {NoStop}%
\bibitem [{\citenamefont {{Lehar}}\ \emph {et~al.}(1992)\citenamefont
  {{Lehar}}, \citenamefont {{Hewitt}}, \citenamefont {{Roberts}},\ and\
  \citenamefont {{Burke}}}]{Lehar1992}%
  \BibitemOpen
  \bibfield  {author} {\bibinfo {author} {\bibfnamefont {J.}~\bibnamefont
  {{Lehar}}}, \bibinfo {author} {\bibfnamefont {J.~N.}\ \bibnamefont
  {{Hewitt}}}, \bibinfo {author} {\bibfnamefont {D.~H.}\ \bibnamefont
  {{Roberts}}}, \ and\ \bibinfo {author} {\bibfnamefont {B.~F.}\ \bibnamefont
  {{Burke}}},\ }\href {\doibase 10.1086/170887} {\bibfield  {journal} {\bibinfo
   {journal} {\apj}\ }\textbf {\bibinfo {volume} {384}},\ \bibinfo {pages}
  {453} (\bibinfo {year} {1992})}\BibitemShut {NoStop}%
\bibitem [{\citenamefont {{Edelson}}\ and\ \citenamefont
  {{Krolik}}(1988)}]{Edelson1988}%
  \BibitemOpen
  \bibfield  {author} {\bibinfo {author} {\bibfnamefont {R.~A.}\ \bibnamefont
  {{Edelson}}}\ and\ \bibinfo {author} {\bibfnamefont {J.~H.}\ \bibnamefont
  {{Krolik}}},\ }\href {\doibase 10.1086/166773} {\bibfield  {journal}
  {\bibinfo  {journal} {\apj}\ }\textbf {\bibinfo {volume} {333}},\ \bibinfo
  {pages} {646} (\bibinfo {year} {1988})}\BibitemShut {NoStop}%
\bibitem [{\citenamefont {{Rasmussen}}\ and\ \citenamefont
  {{Williams}}(2006)}]{Rasmussen2006}%
  \BibitemOpen
  \bibfield  {author} {\bibinfo {author} {\bibfnamefont {C.~E.}\ \bibnamefont
  {{Rasmussen}}}\ and\ \bibinfo {author} {\bibfnamefont {C.~K.~I.}\
  \bibnamefont {{Williams}}},\ }\href@noop {} {\emph {\bibinfo {title}
  {{Gaussian Processes for Machine Learning}}}}\ (\bibinfo  {publisher} {MIT
  Press},\ \bibinfo {year} {2006})\BibitemShut {NoStop}%
\bibitem [{\citenamefont {{Wilkins}}\ \emph {et~al.}(2021)\citenamefont
  {{Wilkins}}, \citenamefont {{Gallo}}, \citenamefont {{Costantini}},
  \citenamefont {{Brandt}},\ and\ \citenamefont {{Blandford}}}]{Wilkins2021}%
  \BibitemOpen
  \bibfield  {author} {\bibinfo {author} {\bibfnamefont {D.~R.}\ \bibnamefont
  {{Wilkins}}}, \bibinfo {author} {\bibfnamefont {L.~C.}\ \bibnamefont
  {{Gallo}}}, \bibinfo {author} {\bibfnamefont {E.}~\bibnamefont
  {{Costantini}}}, \bibinfo {author} {\bibfnamefont {W.~N.}\ \bibnamefont
  {{Brandt}}}, \ and\ \bibinfo {author} {\bibfnamefont {R.~D.}\ \bibnamefont
  {{Blandford}}},\ }\href {\doibase 10.1038/s41586-021-03667-0} {\bibfield
  {journal} {\bibinfo  {journal} {\nat}\ }\textbf {\bibinfo {volume} {595}},\
  \bibinfo {pages} {657} (\bibinfo {year} {2021})},\ \Eprint
  {http://arxiv.org/abs/2107.13555} {arXiv:2107.13555} \BibitemShut {NoStop}%
\bibitem [{\citenamefont {{Gralla}}\ \emph {et~al.}(2020)\citenamefont
  {{Gralla}}, \citenamefont {{Lupsasca}},\ and\ \citenamefont
  {{Marrone}}}]{GLM2020}%
  \BibitemOpen
  \bibfield  {author} {\bibinfo {author} {\bibfnamefont {S.~E.}\ \bibnamefont
  {{Gralla}}}, \bibinfo {author} {\bibfnamefont {A.}~\bibnamefont
  {{Lupsasca}}}, \ and\ \bibinfo {author} {\bibfnamefont {D.~P.}\ \bibnamefont
  {{Marrone}}},\ }\href {\doibase 10.1103/PhysRevD.102.124004} {\bibfield
  {journal} {\bibinfo  {journal} {\prd}\ }\textbf {\bibinfo {volume} {102}},\
  \bibinfo {eid} {124004} (\bibinfo {year} {2020})},\ \Eprint
  {http://arxiv.org/abs/2008.03879} {arXiv:2008.03879} \BibitemShut {NoStop}%
\bibitem [{\citenamefont {{Chael}}\ \emph {et~al.}(2021)\citenamefont
  {{Chael}}, \citenamefont {{Johnson}},\ and\ \citenamefont
  {{Lupsasca}}}]{Chael2021}%
  \BibitemOpen
  \bibfield  {author} {\bibinfo {author} {\bibfnamefont {A.}~\bibnamefont
  {{Chael}}}, \bibinfo {author} {\bibfnamefont {M.~D.}\ \bibnamefont
  {{Johnson}}}, \ and\ \bibinfo {author} {\bibfnamefont {A.}~\bibnamefont
  {{Lupsasca}}},\ }\href {\doibase 10.3847/1538-4357/ac09ee} {\bibfield
  {journal} {\bibinfo  {journal} {\apj}\ }\textbf {\bibinfo {volume} {918}},\
  \bibinfo {eid} {6} (\bibinfo {year} {2021})},\ \Eprint
  {http://arxiv.org/abs/2106.00683} {arXiv:2106.00683} \BibitemShut {NoStop}%
\bibitem [{\citenamefont {{Paugnat}}\ \emph {et~al.}(2022)\citenamefont
  {{Paugnat}}, \citenamefont {{Lupsasca}}, \citenamefont {{Vincent}},\ and\
  \citenamefont {{Wielgus}}}]{Paugnat2022}%
  \BibitemOpen
  \bibfield  {author} {\bibinfo {author} {\bibfnamefont {H.}~\bibnamefont
  {{Paugnat}}}, \bibinfo {author} {\bibfnamefont {A.}~\bibnamefont
  {{Lupsasca}}}, \bibinfo {author} {\bibfnamefont {F.~H.}\ \bibnamefont
  {{Vincent}}}, \ and\ \bibinfo {author} {\bibfnamefont {M.}~\bibnamefont
  {{Wielgus}}},\ }\href {\doibase 10.1051/0004-6361/202244216} {\bibfield
  {journal} {\bibinfo  {journal} {\aap}\ }\textbf {\bibinfo {volume} {668}},\
  \bibinfo {eid} {A11} (\bibinfo {year} {2022})},\ \Eprint
  {http://arxiv.org/abs/2206.02781} {arXiv:2206.02781} \BibitemShut {NoStop}%
\bibitem [{\citenamefont {{Cunningham}}(1975)}]{Cunningham1975}%
  \BibitemOpen
  \bibfield  {author} {\bibinfo {author} {\bibfnamefont {C.~T.}\ \bibnamefont
  {{Cunningham}}},\ }\href {\doibase 10.1086/154033} {\bibfield  {journal}
  {\bibinfo  {journal} {\apj}\ }\textbf {\bibinfo {volume} {202}},\ \bibinfo
  {pages} {788} (\bibinfo {year} {1975})}\BibitemShut {NoStop}%
\bibitem [{\citenamefont {{Lindgren}}\ \emph {et~al.}(2011)\citenamefont
  {{Lindgren}}, \citenamefont {{Rue}},\ and\ \citenamefont
  {{Lindstr\"om}}}]{Lindgren2011}%
  \BibitemOpen
  \bibfield  {author} {\bibinfo {author} {\bibfnamefont {F.}~\bibnamefont
  {{Lindgren}}}, \bibinfo {author} {\bibfnamefont {H.~v.}\ \bibnamefont
  {{Rue}}}, \ and\ \bibinfo {author} {\bibfnamefont {J.}~\bibnamefont
  {{Lindstr\"om}}},\ }\href {\doibase 10.1111/j.1467-9868.2011.00777.x}
  {\bibfield  {journal} {\bibinfo  {journal} {Journal of the Royal Statistical
  Society: Series B (Statistical Methodology)}\ }\textbf {\bibinfo {volume}
  {73}},\ \bibinfo {pages} {423} (\bibinfo {year} {2011})},\ \Eprint
  {http://arxiv.org/abs/https://rss.onlinelibrary.wiley.com/doi/pdf/10.1111/j.1467-9868.2011.00777.x}
  {https://rss.onlinelibrary.wiley.com/doi/pdf/10.1111/j.1467-9868.2011.00777.x}
  \BibitemShut {NoStop}%
\bibitem [{\citenamefont {{Fuglstad}}\ \emph {et~al.}(2015)\citenamefont
  {{Fuglstad}}, \citenamefont {{Lindgren}}, \citenamefont {{Simpson}},\ and\
  \citenamefont {{Rue}}}]{Fuglstad2013}%
  \BibitemOpen
  \bibfield  {author} {\bibinfo {author} {\bibfnamefont {G.-A.}\ \bibnamefont
  {{Fuglstad}}}, \bibinfo {author} {\bibfnamefont {F.}~\bibnamefont
  {{Lindgren}}}, \bibinfo {author} {\bibfnamefont {D.}~\bibnamefont
  {{Simpson}}}, \ and\ \bibinfo {author} {\bibfnamefont {H.}~\bibnamefont
  {{Rue}}},\ }\href {\doibase 10.2307/24311007} {\bibfield  {journal} {\bibinfo
   {journal} {Statistica Sinica}\ }\textbf {\bibinfo {volume} {25}},\ \bibinfo
  {pages} {115} (\bibinfo {year} {2015})},\ \Eprint
  {http://arxiv.org/abs/1304.6949} {arXiv:1304.6949} \BibitemShut {NoStop}%
\end{thebibliography}%

\end{document}